\documentclass[12pt]{article}
\usepackage{epsf}
\usepackage{amsmath,amssymb}
\usepackage{latexsym}
\usepackage{wrapfig}
\usepackage{latexsym}
\usepackage{comment}
\usepackage{color}
\makeatletter
   
   \@addtoreset{equation}{section}
\makeatother

\begin{document}
\baselineskip 0.7cm

\begin{titlepage}

\setcounter{page}{0}

\renewcommand{\thefootnote}{\fnsymbol{footnote}}

\begin{flushright}
YITP-13-112\\
\end{flushright}

\vskip 1.35cm

\begin{center}
{\Large \bf 
Supersymmetric Gauge Theories on a Squashed Four-sphere
}

\vskip 1.2cm 

{\normalsize
Tomoki Nosaka\footnote{nosaka(at)yukawa.kyoto-u.ac.jp} and Seiji Terashima\footnote{terasima(at)yukawa.kyoto-u.ac.jp}
}

\vskip 0.8cm

{ \it
Yukawa Institute for Theoretical Physics, Kyoto University, Kyoto 606-8502, Japan
}

\end{center}

\vspace{12mm}

\centerline{{\bf Abstract}}

\vskip 0.5cm 
We define a squashed four-sphere by a dimensional reduction 
of a twisted $S^4\times S^1$, and construct explicitly 
a supersymmetric Yang-Mills action on it.
The action includes a non-trivial 
dilaton factor and a theta term with a non-constant theta.
The partition function of this theory is calculated using 
the localization technique.
The resulting partition function can be
written in the form consistent with
the AGT relation due to the non-constant theta term.
The parameter $b$ which characterizes the partition function
in this form is not restricted to be real
for the squashed four-sphere.

\end{titlepage}

\newpage
\tableofcontents
\vskip 1.2cm

\section{Introduction and Summary}

Supersymmetric gauge theories on curved compact manifolds are
investigated intensively after 
the theories on the round four-sphere was considered in \cite{Pestun:2007rz}.
One reason for the interests
is that one can calculate exactly some supersymmetric quantities, 
such as the partition function and the expectation value of a Wilson
loop, using the localization technique.
In particular, the theories on four dimensional manifolds 
will be important for the understanding of the non-perturbative
dynamics of the gauge theories related to QCD and beyond the standard model.
The 4d theories are also relevant 
to the check of the relations between 4d gauge theories and 2d
conformal field theories, which is called the AGT relations \cite{Alday:2009aq}.

Despite these obvious importance, 
such computations
have been performed for the 4d supersymmetric gauge theories
only on a 4d ellipsoid \cite{Hama:2012bg}, which includes the round
four-sphere,
and on $S^3 \times S^1$ which gives the index \cite{Gadde:2011ia,Imamura:2011uw}.\footnote{
Though the exact calculation of the path integral was not performed, a kind of a continuous deformation of the round four-sphere, neither included in the ellipsoids nor in the squashed four sphere we define in this paper, was also considered in \cite{Liu:2012bi}.
}
Therefore, it will be important to find 
other manifolds on which we can define supersymmetric gauge theories
and to compute exactly some quantities of them.
There are many possibilities to find such manifolds.
However, 
one of the simplest constructions might be 
a ``squashed four-sphere''.
In \cite{Imamura:2012bm} it was shown that squashed $(2n-1)$-spheres,
where $n$ is a positive integer, can
be constructed by a dimensional reduction of $S^{2n-1}\times S^1$, where
the reduced circular direction is 
a linear combination of the $S^1$ direction and 
an isometry on the $(2n-1)$-sphere.
This procedure will be called the twisted dimensional reduction.
When the same procedure is applied to $S^{2n}\times S^1$, however, 
one might expect that some singularity may
appear since any isometry on the $(2n)$-sphere has fixed points, 
i.e. the north pole and the south pole.

In this paper, we show explicitly that when we apply the 
above process to $S^4\times S^1$, 
we obtain a non-singular manifold which we will call a squashed
four-sphere, which has two deformation parameters
$\epsilon_1, \epsilon_2$.
It can be shown that the supersymmetries on it are similar to those on
untwisted $S^4\times S^1$ \cite{Kim:2012gu,Terashima:2012ra}.
However, some of the supersymmetries 
are projected out for the compatibility with the twisting.

Since a supersymmetric Yang-Mills action on $S^4 \times S^1 $
may not be possible \cite{Terashima:2012ra},
we need to study whether there
exist any supersymmetric Yang-Mills action even if the squashed four-sphere
 is well defined.
Indeed,  for $\epsilon_1=-\epsilon_2$,
we succeed to construct explicitly  
a supersymmetric Yang-Mills action on
the squashed four-sphere,
which includes a theta term with a non-constant theta.
Then, the instanton factor near the north pole and the south pole of 
the squashed four-sphere becomes
\begin{align}
\tau_{\text{eff}}=\frac{\theta_0}{2\pi}+\frac{4\pi i (1+(\epsilon_1)^2 l^2)}{{g_{\text{YM}}}^2},
\end{align}
due to the non-constant theta term.
Then we calculate the partition function by the localization technique.
The result is written as
$ Z(\tau_{\text{eff}},b, \mu)$
where $\mu$ is an effective mass parameter for the hypermultiplet and
\begin{align}
b=\sqrt{\frac{1-i\ell\epsilon_1}{1-i\ell\epsilon_2}}.
\end{align}
%
%

%
We find that this $Z(\tau,b,\mu)$ is the same function which appears in \cite{Hama:2012bg} for the partition function 
of the theory on the ellipsoid and consistent with 
the AGT relation.
Here it is non-trivial that
only the three parameters out of 
the various parameters of the theory
appear in 
the partition function 
$Z(\tau,b,\mu)$ due to, for example, the non-constant theta term.
It should be noted that 
in our case, because $\epsilon_1$ and $\epsilon_2$ can be taken 
to be arbitrary real values, 
one can realize
arbitrary value of $b^2$ in $\mathbb{C}\backslash\mathbb{R}_+$.

For the 4d ellipsoid, 
the partition function obtained in \cite{Hama:2012bg} 
has a real parameter $b$, 
which is the square root of the ratio between the length of the major
semi-axis and that of the minor semi-axis.  
In this case, however, one cannot make $b$ complex because
the metric is complex for generic $b$
or the manifold is non-compact for pure imaginary $b$
which implies that the path integral will be IR-divergent.
Remarkably, in the case of the squashed sphere, in contrast to the case of
the ellipsoid, 
we can take both $\epsilon_{1}$  and $\epsilon_{2}$ 
pure imaginary 
such that $|\epsilon_i|<\frac{1}{\ell}$.
Including this region, 
the parameter $b$ can take any value in $\mathbb{R}_+$ also.
The partition function with the complex $b$
could be interesting implication for the AGT relation
because naive guess for the CFT counter part
will have complex central charge \cite{Harlow:2011ny}.
We will leave this problem in future investigations.

We note that 
there is an analogy 
with the theories on the deformed three dimensional spheres
for the dependence of the deformation parameters.
Actually, the three dimensional ellipsoid
\cite{Hama:2011ea} and the squashed three-sphere \cite{Imamura:2011wg}
were considered and the 
the partition functions are in the same form for the both cases
with one parameter $b$.
However, while in the former case the parameter $b$ is real, having
similar geometrical meaning as that in the 4d case,
in the latter case it is a complex number.
For the three-dimensional cases, 
it were explained in \cite{Closset:2012ru,Alday:2013lba} why these two deformations 
with the different geometrical origins
give the partition functions in a same form.
For our four dimensional cases,
we do no have a clear explanations
why the partition functions take in the same form\footnote{
A partial explanation is that because 
the partition function can be computed at the localized points
where the Nekrasov's omega deformations are expected to be
the only possible relevant supersymmetric deformations
\cite{Hosomichisprivatecommunication}.
In our case, the weights in the classical action at north pole and 
the south pole are modified, however, the partition function is 
in the same form. The AGT relation and 
the M5-branes are expected to be a possible origin of this property.}
although there are similar works \cite{Closset:2013vra,Pan}.
This is because 
we consider a deformation of ${\cal N}=2$ supersymmetric
gauge field theories and the theory on the squashed four-sphere
is coupled to the non-trivial ``dilaton'' as we will explain.
The investigations for these lines will be also interesting.

From our results, we expect that there are a few different 
supersymmetric partition functions on compact four
dimensional manifolds.
It would be interesting to have other four dimensional manifolds
on which the supersymmetric partition function is different from
the one considered in the paper.
One of the possible directions 
is study of supersymmetric theories on a manifold with boundaries.
Indeed, such examples have been considered for 
two and three dimensional manifolds with boundaries in
\cite{SuTe, HoOk, HoRo}.
We hope to report some results in this direction
in near future.

The rest of this paper is organized as follows.
In Section 2 we define a squashed four-sphere.
To obtain the supersymmetry on it, in Section 3 we review the supersymmetry on $S^4\times S^1$.
In section 4 we confirm that there are supersymmetries on $S^4\times S^1$ which are compatible with the twist of the periodicity.
After the dimensional reduction, they turn to the supersymmetry on the squashed four-sphere.
A supersymmetric Yang-Mills action on the squashed four-sphere is also constructed in section 5.
In section 6 we calculate the partition function of the theory defined in the previous sections.

\section{Squashed four-spheres by the twisted dimensional reduction}

In this section, we will define squashed four-spheres.
In \cite{Imamura:2012bm} the squashed $(2n-1)$-spheres were constructed 
by a dimensional reduction of $S^{2n-1}\times S^1$, 
where the reduced circular direction
is a linear combination of the $S^1$ direction
and a $U(1)$ isometry direction on the $(2n-1)$-sphere.
In this paper we call this process as twisted dimensional reduction.

We apply a similar dimensional reduction to $S^4\times S^1$.
Unlike $S^{2n-1}\times S^1$, the $U(1)$ directions on $S^4$
shrinks to a point
at the North pole and the South pole, and thus one might expect
that the twisted dimensional reduction would produce a singular
manifold.
As we will see below, however, the resulting manifold is actually
non-singular, and we call it a squashed four-sphere.
Since the isometry group on the round 
four-sphere, $SO(5)$, contains two $U(1)$'s which
are consistent with the supersymmetry as we will see later,
the squashed four-spheres are two parameter deformations of the round
four-sphere.

First, the metric on $S^4\times S^1$ is
\begin{align}
\mathrm{d}s^2=\mathrm{d}s_{s^4}^4+\mathrm{d}t^2\label{5dmetric},
\end{align}
where $t$ is the coordinate of 
the $S^1$ direction,
with $t\sim t+2\pi\beta$.\footnote{
Here we used the symbol $\sim$ as an identification.} 
Here $\beta$ is the radius of the circle.
The metric on the round four-sphere is given by
\begin{align}
\mathrm{d}s_{S^4}^2=\ell^2\{\mathrm{d}\theta^2+\mathrm{sin}\theta^2(\mathrm{d}\phi^2+\mathrm{cos}^2\phi\mathrm{d}\alpha_1^2+\mathrm{sin}^2\phi\mathrm{d}\alpha_2^2)\},
\end{align}
where $\ell$ is the radius of the four-sphere.
The domain of the coordinates are $0\le\theta\le\pi$,
$0\le\phi\le\frac{\pi}{2}$, $0\le\alpha_1\le 2\pi$ and 
$0\le\alpha_2\le 2\pi$, 
and $\alpha_{1,2}$ are periodic coordinates.

To obtain a squashed four-sphere, we first 
define the twisted $S^4 \times S^1$ which has 
the same metric $ds^2$ as the $S^4 \times S^1$, but 
the periodicities in
the three circular directions $(\alpha_1,\alpha_2,t)$ are given as
\begin{align}
\begin{pmatrix}
\alpha_1\\
\alpha_2\\
t
\end{pmatrix}
\sim
\begin{pmatrix}
\alpha_1-2\pi\beta\epsilon_1n_3+2\pi n_1\\
\alpha_2-2\pi\beta\epsilon_2n_3+2\pi n_2\\
t+2\pi\beta n_3\label{twistedpbc}
\end{pmatrix}
\,\,\,\,\,\,(n_1,n_2,n_3 \in \mathbb{Z}),
\end{align}
where we introduced two real parameter $\epsilon_1$ and $\epsilon_2$.
Introducing new coordinates defined as
\begin{align}
\alpha_1^\prime\equiv\alpha_1+\epsilon_1t,\,\,\,\,\,\,\alpha_2^\prime\equiv\alpha_2+\epsilon_2t,\,\,\,\,\,\,t^\prime\equiv t,
\end{align}
(\ref{twistedpbc}) is written in an ``untwisted'' form:
\begin{align}
\begin{pmatrix}
\alpha_1^\prime\\
\alpha_2^\prime\\
t^\prime
\end{pmatrix}
\sim
\begin{pmatrix}
\alpha_1^\prime+2\pi n_1\\
\alpha_2^\prime+2\pi n_2\\
t^\prime+2\pi\beta n_3\label{twistedper}
\end{pmatrix}.
\end{align}
Now it is possible to reduce the new circular direction denoted by
$t^\prime$, by taking the limit of $\beta \rightarrow 0$.
To obtain the metric of resulting four dimensional manifold, we rewrite
the metric on the twisted $S^4\times S^1$ (\ref{5dmetric}) in the new
coordinates as
\begin{align}
\mathrm{d}s^2&=\ell^2\{\mathrm{d}\theta^2
  +\mathrm{sin}^2\theta(\mathrm{d}\phi^2
    +\mathrm{cos}^2\phi\mathrm({d}\alpha_1^\prime-\epsilon_1\mathrm{d}t^\prime)^2
    +\mathrm{sin}^2\phi\mathrm({d}\alpha_2^\prime-\epsilon_2\mathrm{d}t^\prime)^2)\}
+\mathrm{d}{t^\prime}^2\nonumber\\
&=\mathrm{d}s_4^2(\epsilon_1,\epsilon_2)+h^2(\mathrm{d}{t^\prime}+u)^2,\label{twisted5d}
\end{align}
where
\begin{align}
\mathrm{d}s_4^2(\epsilon_1,\epsilon_2)=\ell^2\left\{\mathrm{d}^2\theta+\mathrm{sin}^2\theta\left(\mathrm{d}\phi^2+\mathrm{cos}^2\phi\mathrm{d}{\alpha_1^\prime}^2+\mathrm{sin}^2\phi\mathrm{d}{\alpha_2^\prime}^2-\frac{h^2}{\ell^2\mathrm{sin}^2\theta}u^2\right)\right\}\label{ds2sqs4},
\end{align}
with
\begin{align}
h&=\sqrt{1+\ell^2\mathrm{sin}^2\theta(\epsilon_1^2\mathrm{cos}^2\phi+\epsilon_2^2\mathrm{sin}^2\phi)},\nonumber\\
u&=u_\mu\mathrm{d}x^\mu=-\frac{\ell^2\mathrm{sin}^2\theta}{h^2}(\epsilon_1\mathrm{cos}^2\phi\mathrm{d}\alpha_1^\prime+\epsilon_2\mathrm{sin}^2\phi\mathrm{d}\alpha_2^\prime).
\end{align}

To see that $\mathrm{d}s^2_4$ is the background metric of the four
dimensional theory obtained by the dimensional reduction of the five
dimensional theory on twisted $S^4\times S^1$, we rewrite the standard kinetic
term of a five dimensional scalar field in the twisted coordinate
$(x^\prime,t^\prime)$:
\begin{align}
&\int\mathrm{d}^4x\mathrm{d}t\sqrt{g(x,t)}g^{mn}(x,t)\partial_m\Phi\partial_n\Phi\nonumber\\
=&\int\mathrm{d}t^\prime\mathrm{d}^4x^\prime\sqrt{G}h(G^{\mu^\prime\nu^\prime}\partial_{\mu^\prime}\Phi\partial_{\nu^\prime}\Phi+2G^{\mu^\prime\nu^\prime}u_{\mu^\prime}\partial_{\nu^\prime}\Phi\partial_{t^\prime}\Phi+(\partial_{t^\prime}\Phi)^2).
\end{align}
where $g_{mn}$ is the five dimensional metric on the twisted 
$S^4\times S^1$, i.e.  the metric defined by (\ref{5dmetric}), 
in $(x,t)$ coordinates
and 
$G_{\mu' \nu'}$ is the metric defined by (\ref{ds2sqs4}).
After taking the limit of $\beta\rightarrow 0$, all the modes with
non-zero Kaluza-Klein momentum decouple, and the action of Kaluza-Klein
zero modes, which can be considered as four dimensional fields, is
\begin{align}
\int\mathrm{d}^4x^\prime\sqrt{G}hG^{\mu^\prime\nu^\prime}\partial_{\mu^\prime}\Phi\partial_{\nu^\prime}\Phi.
\end{align}
This implies that the four dimensional theory obtained by the
dimensional reduction have background metric of the squashed four-sphere
$G_{\mu^\prime\nu^\prime}$.
Note that this theory also has the non-constant dilaton background $h$.
By considering a vector field instead of the scalar field,
one finds that 
the theory have the graviphoton background $u$ also.

The four dimensional manifold with 
the metric (\ref{ds2sqs4}) is actually non-singular and thus well-defined.
Indeed, the last term in (\ref{ds2sqs4}) behaves as $\ell^4\mathrm{sin}^4\theta\epsilon_i^2\mathrm{d}\alpha_j^2$, which vanishes appropriately at the north pole and the south pole.
This metric is the one on the round $S^4$ deformed by two parameters in a similar way to obtain a squashed $(2n-1)$-sphere from $S^{2n-1}\times S^1$ in \cite{Imamura:2012bm}.
In this paper we call the manifold with the 
metric (\ref{ds2sqs4}) as a squashed four-sphere.

So far we have assumed that both $\epsilon_i$'s are real. 
Here let us take $\epsilon_i$'s pure imaginary formally in the metric
(\ref{ds2sqs4}).
Replacing $\epsilon_i$ with $i {e}_i $, one obtains
\begin{align}
\mathrm{d}s_4^2(i{e}_1,i{e}_2)=\ell^2\left\{\mathrm{d}^2\theta+\mathrm{sin}^2\theta\left(\mathrm{d}\phi^2+\mathrm{cos}^2\phi\mathrm{d}{\alpha_1^\prime}^2+\mathrm{sin}^2\phi\mathrm{d}{\alpha_2^\prime}^2+\frac{\tilde{h}^2}{\ell^2\mathrm{sin}^2\theta}u^2\right)\right\}
\end{align}
with $u({e}_1,{e}_2)$ and
\begin{align}
\tilde{h}({e}_1,{e}_2)&=\sqrt{1-\ell^2\mathrm{sin}^2\theta({e}_1^2\mathrm{cos}^2\phi+{e}_2^2\mathrm{sin}^2\phi)}.
\end{align}
In this case the metric becomes singular at $\theta\sim \frac{\pi}{2}$ if $\displaystyle{\max_i\{{e}_i^2\}\ge \ell^2}$.
However, for ${e}_i^2<\ell^2$ one still have a non-singular manifold.
Therefore, 
we have non-singular manifolds 
for $\epsilon_i$'s being both real or 
$\epsilon_i$'s being both imaginary with $(i \epsilon_i)^2 < \ell^2$.

\section{Supersymmetric gauge theories on $S^4\times S^1$}

The supersymmetry transformation law on the squashed four-sphere 
can be obtained if we have the one on the twisted $S^4 \times S^1$
because the supersymmetry transformation are closed 
even if the fields are restricted to the Kaluza-Klein zero-modes.
Because the difference between 
the $S^4\times S^1$ and the twisted $S^4\times S^1$
is just in the coordinate identifications,
the supersymmetry transformation on the twisted $S^4\times S^1$
can be obtained from the one on the $S^4\times S^1$ 
as we will show later.
Thus, in this section, we review the supersymmetry on the $S^4\times S^1$ 
which were constructed in
\cite{Kim:2012gu,Terashima:2012ra}.


Here we use the convention used in \cite{HST} \cite{Terashima:2012ra} 
unless otherwise stated.
We use Greek indices ($\mu,\nu,\cdots=1,2,3,4$) for the directions along the four-sphere and $t$ or $5$ for the circular direction.
To represent all the five directions, 
we use $m,n,\cdots=1,2,3,4,5$.
When an index denote the corresponding local Lorentz index, 
a symbol ``hat'' is added on it (for examples, $\hat{\mu}$ and $\hat{m}$).

\subsection*{Vectormultiplet}

The vectormultiplet 
in the five dimensional ${\cal N}=1$ supersymmetry
consists of a connection 1-form $A_m$, a (Hermite) scalar boson $\sigma$, 
two spinor fermions $\lambda_I$ and three auxiliary scalar bosons $D_{IJ}$.
Here $I,J=1,2$ and there is an $SU(2)_R$ symmetry acting on them.
The fermions $\lambda_I$ transform as a doublet under this $SU(2)_R$.
The auxiliary fields
$D_{IJ}$ are symmetric with respect to the two indices and transforms as
a triplet under the $SU(2)_R$.

The supersymmetry transformation law of the vectormultiplet is given by
\begin{align}
\delta_\xi A_m&=i\epsilon^{IJ}\xi_I\Gamma_m\lambda_J,\nonumber\\
\delta_\xi\sigma&=i\epsilon^{IJ}\xi_I\lambda_J,\nonumber\\
\delta_\xi\lambda_I&=-\frac{1}{2}\Gamma^{mn}\xi_IF_{mn}+\Gamma^m\xi_ID_m\sigma+\xi_JD_{KI}\epsilon^{JK}+2t_I^{\phantom{I}J}\Gamma_5\xi_J\sigma,\nonumber\\
\delta_\xi D_{IJ}&=-i\xi_I\Gamma^mD_m\lambda_J+[\sigma,\xi_I\lambda_J]+i\xi_Kt_{IJ}\Gamma_5\lambda^K+(I\leftrightarrow J),\label{deltalambda}
\end{align}
where $\xi_I$ are the spinors satisfying following Killing Spinor equations
\begin{align}
D_\mu\xi_I=t_I^{\phantom{I}J}\Gamma_\mu\Gamma_5\xi_J,\,\,\,\,\,\, D_5\xi_I=0.\label{KSeq}
\end{align}
We also impose the twisted $SU(2)$ Majorana condition
\cite{Terashima:2012ra}, 
$\xi_I^\dagger=-\epsilon^{IJ}\,^t\!\xi_JC\Gamma_5$, where $C$ is the charge conjugation matrix.
Here $t_I^{\phantom{I}J}$ is a constant matrix which can be chosen as
\begin{align}
t_I^{\phantom{I}J}=\frac{1}{2\ell}(\sigma_3)_I^{\phantom{I}J}.
\end{align}

The equations (\ref{KSeq}) can be explicitly solved as
\begin{align}
\xi_I=\sqrt{f}\left(1-\frac{(-1)^I}{2\ell}\Gamma^{\hat{\mu}}x_\mu\Gamma_5\right)\psi_I,
\label{KSexplicit}
\end{align}
with a pair of constant spinors $(\psi_1,\psi_2)$ related to each other so that $\xi_I$ satisfy the twisted $SU(2)$ Majorana condition.
Here we have chosen the coordinate $x_\mu$ as a conformal basis defined by
\begin{align}
x_1&=2 \ell \, \mathrm{tan}\frac{\theta}{2} \, \mathrm{cos}\phi \, \mathrm{cos}\alpha_1,\nonumber\\
x_2&=2\ell\mathrm{tan}\frac{\theta}{2}\mathrm{cos}\phi\mathrm{sin}\alpha_1,\nonumber\\
x_3&=2\ell\mathrm{tan}\frac{\theta}{2}\mathrm{sin}\phi\mathrm{cos}\alpha_2,\nonumber\\
x_4&=2\ell\mathrm{tan}\frac{\theta}{2}\mathrm{sin}\phi\mathrm{sin}\alpha_2,\label{conformalbasis}
\end{align}
with which the metric on the four-sphere is simply $\mathrm{d}s_{S^4}^2=f^2\mathrm{d}x_\mu\mathrm{d}x_\mu$, where $f=\left(1+\frac{x_\mu^2}{4\ell^2}\right)^{-1}$.

{}From (\ref{deltalambda}) one finds that a commutator of two
supersymmetry transformations close into the sum of a translation
generated by a linear combination of 
Killing vectors, a $SU(2)_R$ rotation, a local Lorentz rotation and 
a gauge transformation:
\begin{align}
[\delta_\xi,\delta_\eta]A_m&=-iv^nF_{nm}+D_m\gamma,\nonumber\\
[\delta_\xi,\delta_\eta]\sigma&=-iv^nD_n\sigma,\nonumber\\
[\delta_\xi,\delta_\eta]\lambda_I&=-iv^nD_n\lambda_I+i[\gamma,\lambda_I]+R_I^{\phantom{I}J}\lambda_J+\frac{1}{4}\Theta^{\hat{a}\hat{b}}\Gamma^{\hat{a}\hat{b}}\lambda_I,\nonumber\\
[\delta_\xi,\delta_\eta]D_{IJ}&=-iv^nD_nD_{IJ}+i[\gamma,D_{IJ}]+R_I^{\hat{I}K}D_{KJ}+R_J^{\phantom{J}K}D_{IK},\label{vecalg}
\end{align}
where
\begin{align}
v^m&=2\epsilon^{IJ}\xi_I\Gamma^m\eta_J,\,\,\,\,\,\,\gamma=-2i\epsilon^{IJ}\xi_I\eta_J\sigma,\,\,\,\,\,\,R_{IJ}=4i\epsilon^{KL}\xi_K\Gamma_5t_{IJ}\eta_L,\nonumber\\
\Theta^{\hat{a}\hat{b}}&=-2it^{IJ}(\xi_I\Gamma_5\Gamma^{\hat{a}\hat{b}}\eta_J-\eta_I\Gamma_5\Gamma^{\hat{a}\hat{b}}\xi_J).
\end{align}

For $S^4 \times S^1$,
it may be impossible to construct
a supersymmetric Yang-Mills action which becomes
the supersymmetric Yang-Mills action on the round four-sphere
in the limit of $\beta\rightarrow 0$ \cite{Terashima:2012ra}.
We will, however, construct a supersymmetric Yang-Mills action 
on the squashed four-sphere in section 4.

\subsection*{Hypermultiplets}
The hypermultiplet consists of two bosonic scalars $q_I$, a fermionic spinor $\psi$ and two bosonic auxiliary scalars ${\cal F}_{I^\prime}$, where $I^\prime=1,2$.
The supersymmetry transformation law of hypermultiplet is given as
\begin{align}
\delta_\xi q_I&=-2i\xi_I\psi,\nonumber\\
\delta_\xi\psi&=\epsilon^{IJ}\Gamma^m\xi_ID_mq_J+i\epsilon^{IJ}\xi_I\sigma q_J-2t^{IJ}\Gamma_5\xi_Iq_J+\epsilon^{I^\prime J^\prime}\check{\xi}_{I^\prime}{\cal F}_{J^\prime},\nonumber\\
\delta_\xi{\cal F}_{I^\prime}&=2\check{\xi}_{I^\prime}(i\Gamma^mD_m\psi+\sigma\psi+\epsilon^{KL}\lambda_Kq_L),\label{deltahyp}
\end{align}
where $\check{\xi}_{I^\prime}$ is a pair of spinors which satisfy
\begin{align}
\xi_I\check{\xi}_{J^\prime}=0,\,\,\,\,\,\,\epsilon^{I^\prime J^\prime}\check{\xi}_{I^\prime}\check{\xi}_{J^\prime}=\epsilon^{IJ}\xi_I\xi_J.\label{xicheckcond}
\end{align}

Since $\check{\xi}_{I^\prime}$ is related to $\xi_I$, the off-shell construction is possible only with respect to a single supersymmetry $\delta_\xi$, but this is enough for the localization technique to be applicable.
Regarding $\delta_\xi$ as a fermionic operation, 
i.e. $\xi_I$ as a bosonic variable, the square of $\delta_\xi$ is
given by
\begin{align}
\delta^2_\xi q_I&=iv^mD_mq_I-i\gamma q_I-R_I^{\phantom{I}J}q_J,\nonumber\\
\delta^2_\xi\psi&=iv^mD_m\psi-i\gamma\psi-\frac{1}{4}\Theta^{\hat{a}\hat{b}}\Gamma^{\hat{a}\hat{b}}\psi,\nonumber\\
\delta^2_\xi{\cal F}_{I^\prime}&=iv^mD_m{\cal F}_{I^\prime}-i\gamma {\cal F}_{I^\prime}+{R^\prime}_{I^\prime}^{\phantom{I^\prime}J^\prime}{\cal F}_{J^\prime},\label{hypsusy}
\end{align}
where 
\begin{align}
v^m&=\epsilon^{IJ}\xi_I\Gamma^m\xi_J,\,\,\,\,\,\,\gamma=-i\epsilon^{IJ}\xi_I\xi_J\sigma,\,\,\,\,\,\,\Theta^{\hat{a}\hat{b}}=2it^{IJ}\xi_I\Gamma_5\Gamma^{\hat{a}\hat{b}}\xi_J,\nonumber\\
R_{IJ}&=2it_{IJ}\epsilon^{KL}\xi_K\Gamma^5\xi_L,\,\,\,\,\,\,R^\prime_{I^\prime J^\prime}=-2i\check{\xi}_{I^\prime}\Gamma^mD_m\check{\xi}_{J^\prime}.
\end{align}
To show (\ref{hypsusy}), following identities are useful
\begin{align}
\epsilon^{IJ}\xi_I\,^t\!\xi_JC+\epsilon^{I^\prime J^\prime}\check{\xi}_{I^\prime}\,^t\!\check{\xi}_{J^\prime}C=-\frac{1}{2}(\xi_I\xi^I)\cdot 1\!\!1,\,\,\,\,\,\,\check{\xi}_{I^\prime}\Gamma^m\check{\xi}^{I^\prime}=-\xi_I\Gamma^m\xi^I,
\end{align}
where $1\!\!1$ is the identity matrix with spinor indices.
These are derived from (\ref{xicheckcond}).

For the hypermultiplet, one can construct an action on $S^4\times S^1$.
The result is
\begin{align}
S_{\text{hyp}}=&\int\!\mathrm{d}^5x\sqrt{g}\mathrm{tr}\Bigg(\epsilon^{IJ}(D_\mu\bar{q}_ID^\mu q_J-\bar{q}_I\sigma^2q_J)-2(i\bar{\psi}\Gamma^mD_m\psi+\bar{\psi}\sigma\psi)\nonumber\\
&\phantom{\int\!\mathrm{d}^5x\sqrt{g}\mathrm{tr}\Bigg(}-i\bar{q}_ID^{IJ}q_J-4\epsilon^{IJ}\bar{\psi}\lambda_Iq_J-\epsilon^{I^\prime J^\prime}\bar{\cal F}_{I^\prime}{\cal F}_{J^\prime}\nonumber\\
&\phantom{\int\!\mathrm{d}^5x\sqrt{g}\mathrm{tr}\Bigg(}-2t^{IJ}\bar{q}_ID_5q_J+\frac{4}{\ell^2}\epsilon^{IJ}\bar{q}_Iq_J\Bigg).\label{Shypcomp}
\end{align}
Above the massless case have been considered for simplicity.
A mass can be introduced as follows.
First add a $U(1)$ vectormultiplet $(A^{(m)}_m,\sigma^{(m)},\lambda^{(m)}_I,D^{(m)}_{IJ})$, and assign a unit $U(1)$ charge to the hypermultiplet.
Then fix the value of this vectormultiplet to a supersymmetry invariant configuration $\{A_5=m$, others$=0\}$.
As a result mass terms of hypermultiplet is induced by the covariant derivatives.

\section{Supersymmetry on a squashed four-sphere}

In this section,
we will show how
a four dimensional theory on the squashed four-sphere 
is obtained by the dimensional reduction of
the five dimensional
theory on the twisted $S^4\times S^1$.
It will be shown that a quarter of the supersymmetry survives if the
periodicity of the fields in the circle 
is appropriately twisted.

First we consider a basis of the Killing spinors as
\begin{align}
\{\xi\}=\bigoplus_{s_1,s_2=\pm 1}\{\xi\}_{(s_1,s_2)},
\end{align}
where $\{\xi\}_{(s_1,s_2)}$ is the set of killing spinors constructed by $\psi_2$ satisfying $\Gamma^{\hat{1}\hat{2}}\psi_2=is_1\psi_2$ and $\Gamma^{\hat{3}\hat{4}}\psi_2=is_2\psi_2$.
Then we find by an explicit calculation that the killing spinors
belonging to each $\{\xi\}_{(s_1,s_2)}$ satisfy the following identity
\begin{align}
{\cal O}(s_1,s_2)\xi_I&\equiv {\cal L}(\partial_t-p)\xi_I+\tilde{R}(s_1,s_2)_I^{\phantom{I}J}\xi_J\nonumber\\
&=0,\label{defofcalO}
\end{align}
where $p=\epsilon_i\partial_{\alpha_i}$ and
\begin{eqnarray}
\tilde{R}(s_1,s_2)_I^{\phantom{I}J}=-i(\epsilon_1s_1+\epsilon_2s_2)\ell
t_I^{\phantom{I}J}.
\label{rt}
\end{eqnarray}
The action of the 
operator ${\cal O}(s_1,s_2)$ is defined by the r.h.s in the first line.
Here the Lie derivative of a spinor is defined by
\begin{align}
{\cal L}(a)\varphi\equiv \left(a^mD_m-\frac{1}{4}(\nabla_ma_n)\Gamma^{mn}\right)\varphi.
\end{align}
For later convenience, we divide ${\cal O}$ into a derivative
$\partial_t-p^\mu\partial_\mu=\partial_{t^\prime}$ and the other
operator without the derivative, and
rewrite (\ref{defofcalO}) as
\begin{align}
\partial_{t^\prime}\xi_I=-\tilde{\cal O}(s_1,s_2)\xi_I\label{defofcalOtil},
\end{align}
where $\tilde{\cal O}(s_1,s_2)$ is the non-derivative part of 
${\cal O}(s_1,s_2)$, i.e. $\tilde{\cal O}(s_1,s_2)
={\cal O}(s_1,s_2)-\partial_{t^\prime} $.

This equation implies 
that the Killing spinor satisfies 
\begin{align}
\xi(x^\prime,t^\prime+2\pi\beta)=\mathrm{e}^{-2\pi\beta\tilde{\cal
 O}(s_1,s_2)}\xi(x^\prime,t^\prime).
\end{align}
Thus, we should modify the boundary condition of the fields
in order to keep (some part of) supersymmetry on the twisted 
$S^4 \times S^1$.

Now we choose one of the $(s_1,s_2)$ and impose following twisted
periodic boundary condition to the fields as in \cite{Imamura:2011wg}:
\begin{align}
\Phi(x^\prime,t^\prime+2\pi\beta)=\mathrm{e}^{-2\pi\beta\tilde{\cal O}(s_1,s_2)}\Phi(x^\prime,t^\prime).\label{twper}
\end{align}
Indeed, with this boundary condition,
we can see that 
the supersymmetry transformations (\ref{deltalambda}) 
are 
consistent for the theory on the twisted $S^4 \times S^1$ 
with the quarter of the supersymmetry with $\xi_I$ belonging to the $\{\xi\}_{(s_1,s_2)}$.

The Kaluza-Klein decomposition of the fields with the 
twisted boundary condition is
\begin{align}
\Phi(x^\prime,t^\prime)=\sum_{n\in \mathbb{Z}}\mathrm{e}^{\frac{int^\prime}{\beta}}\mathrm{e}^{-t^\prime\tilde{\cal O}(s_1,s_2)}\Phi_n(x^\prime).\label{twistedKKdec}
\end{align}
Therefore, if we take the limit of $\beta\rightarrow 0$, 
an ordinal kinetic term give infinitely large masses in four-dimensional sense to each Kaluza-Klein modes except those with $n=0$.
As a result all the modes with $n\neq 0$ decouple and the fields are
effectively restricted so that they have only the Kaluza-Klein zero
modes which satisfy
\begin{align}
\partial_{t^\prime}\Phi=-\tilde{\cal O}(s_1,s_2)\Phi.\label{twistedrestriction}
\end{align}

The identities (\ref{defofcalOtil}) guarantees that this condition can be kept under the supersymmetry transformation $\delta_\xi$ with $\xi\in\{\xi\}_{(s_1,s_2)}$.\footnote{
By simply substituting the Kaluza-Klein decomposition into the r.h.s of the supersymmetry transformation, one finds that the modes with non-zero Kaluza-Klein momentum also contribute to the Kaluza-Klein zero mode of the l.h.s. through non-linear terms.
However these contribution vanish after renormalizing the modes with $n\neq 0$ as $\Phi_n\rightarrow \frac{\beta}{n}\Phi$ and taking the limit of $\beta\rightarrow 0$.
}
Therefore these supersymmetry is preserved even in the four dimensional
theory on the squashed four-sphere obtained by the dimensional reduction.

The theory on the squashed four-sphere has generically a quarter of the 
supersymmetry of the theory on the round four-sphere.
However, for some special values of the deformation parameters $\epsilon_i$,
supersymmetry enhancements occur. 
For generic $\epsilon_1$ and $\epsilon_2$, 
the conditions 
$\partial_{t^\prime}\Phi=\tilde{\cal O}(s_1,s_2)$ with different
$(s_1,s_2)$s
are distinct.
Therefore one can impose only one of four conditions and as a result the
surviving supersymmetry is a quarter of those in the theory on $S^4 \times S^1$.
On the other hand, if one of $\epsilon_i$ is zero or $\epsilon_1\pm\epsilon_2=0$, two of the four conditions degenerate.
Correspondingly the supersymmetry in four dimension enhances to a half of those in the theory on $S^4 \times S^1$.
Moreover, if $\epsilon_1=\epsilon_2=0$, all the four condition
degenerate and all the supersymmetry
are preserved.
In this case, the four dimensional theory is just the ${\cal N}=2$ theory on the round four-sphere \cite{Pestun:2007rz}.

The supersymmetry transformation law in four dimensional theory 
on the squashed four-sphere is immediately obtained by replacing all the $\partial_{t^\prime}$ in the five dimensional law (\ref{deltalambda}) and (\ref{deltahyp}) with $\tilde{\cal O}(s_1,s_2)$, according to (\ref{twistedrestriction}).
An action of the four dimensional theory also would be obtained from
that of the five dimensional theory on the twisted $S^4\times S^1$ 
by a similar replacement.
Unfortunately, however, no such five dimensional action is known which would reproduce the four dimensional Yang-Mills term.
Therefore the construction of a supersymmetry invariant Yang-Mills
action on a squashed four-sphere is a non-trivial problem.
In the next section we will consider this problem.

\section{Supersymmetry invariant action of the vectormultiplet on a squashed four-sphere}

In this section we will construct an supersymmetry invariant Yang-Mills action on the squashed four-sphere.

As pointed out in \cite{Terashima:2012ra} it is difficult to construct a supersymmetry invariant Yang-Mills action on $S^4\times S^1$.
The reason is as follows.
On the one hand, a four dimensional Yang-Mills action on the four-sphere is obtained from the one on the flat $\mathbb{R}^4$ by conformal transformation.
Through this step the two scalars in vectormultiplet gain masses associated to the curvature.
On the other hand, if one had a five dimensional action and dimensionally reduce the circular direction $t$, $A_t$ turned into a scalar, which must be massless because of the five dimensional gauge symmetry.

It is possible, however, to construct an action on 
the twisted $S^4 \times S^1$ which is both gauge
invariant and supersymmetry invariant if we neglect the massive KK modes.
From this action 
one can obtain an gauge invariant and supersymmetry invariant action on the squashed four-sphere
by the dimensional reduction.\footnote{
Of course, we can construct an gauge and supersymmetry invariant action
on the squashed four-sphere directly using the four dimensional supersymmetry
transformation, however, we guess that a computation for that 
is harder than the five-dimensional construction we employed here.}

Below, we construct such an action.
Hereafter we choose $s_1=s_2=1$
because 
the choice of it is just the
matter of convention and can be absorbed to the definition of
$\epsilon_i$.
Furthermore, for simplicity, 
we restrict the twisting parameters to satisfy $\epsilon_1=-\epsilon_2\equiv\epsilon$.
Even though it might be possible to construct an invariant action for the general $\epsilon_i$, it would require more lengthy and tedious calculation.

We start from the following trial action:
\begin{align}
S_{\text{vec}}^\prime&=\int\!\mathrm{d}^5x\sqrt{g}\mathrm{tr}\Bigg(\frac{1}{2}F_{mn}F^{mn}-(D_m\sigma)(D^m\sigma)
  -\frac{1}{2}D_{IJ}D^{IJ}+i\lambda_I\Gamma^mD_m\lambda^I\nonumber\\
&\phantom{=\int\!\mathrm{d}^5x\sqrt{g}\mathrm{tr}\Bigg(}-\lambda_I[\sigma,\lambda^I]+2A_{t^\prime}t^{IJ}D_{IJ}+\frac{1}{\ell^2}(3A_{t^\prime}^2-2\sigma^2)\Bigg).
\end{align}
This ansatz is motivated by modifying the one constructed in \cite{Terashima:2012ra} so that the twisted dimensional reduction produces appropriate scalar terms (written in $A_{t^\prime}=A_t-p^\mu A_\mu$).
The supersymmetry transformation of this $S_{\text{vec}}^\prime$ is
\begin{align}
\delta_\xi S_{\text{vec}}^\prime&=\int\!\mathrm{d}^5x\sqrt{g}\mathrm{tr}\Bigg(4it^{IJ}\{\xi_I\Gamma^m\lambda_J({\cal O}(1,1)A_m-\xi_I\lambda_J{\cal O}(1,1)\sigma\}\nonumber\\
&\phantom{=\int\!\mathrm{d}^5x\sqrt{g}\mathrm{tr}\Bigg(}-\frac{6i}{\ell^2}p^\mu A_{t^\prime}\xi_I\Gamma_\mu\lambda^I\nonumber\\
&\phantom{=\int\!\mathrm{d}^5x\sqrt{g}\mathrm{tr}\Bigg(}+4ip^\mu t^{IJ}\left\{F_{\mu m}\xi_I\Gamma^m\lambda_J-\xi_I\lambda_JD_\mu\sigma-\frac{1}{2}D_{IJ}\xi_K\Gamma_\mu\lambda^K\right\}\Bigg),
\label{l1}
\end{align}
where we arranged the terms which vanishes as the limit of $\beta\rightarrow 0$ is taken and the fields are restricted as (\ref{twistedrestriction}) into the first line.
To write down the rest of terms in $\delta_\xi S_{vec}^\prime$ (the second line and the third line), we used the explicit form of ${\cal O}$ acting on $A_m$ and $\sigma$
\begin{align}
{\cal O}(1,1)A_m=\partial_{t^\prime}A_m-(\partial_mp^n)A_n,\,\,\,\,\,\,{\cal O}(1,1)\sigma=\partial_{t^\prime}\sigma.
\end{align}
In order to have a supersymmetry invariant action on the squashed four-sphere, 
we need to
add some terms to the trial action $S^\prime_{\text{vec}}$ which compensate the second
line
in (\ref{l1}).
To find such terms, the following three identities will be useful:
\begin{align}
s^{IK}t_K^{\phantom{K}J}+t^{IK}s_K^{\phantom{K}J}=-t^{KL}s_{KL}\epsilon^{IJ},\,\,\,\,\,\,t_I^{\phantom{I}J}\xi_J=F_{(2)}\xi_I,\,\,\,\,\,\, t_I^{\phantom{I}J}\,^t\!\xi_JC=\,^t\!\xi_ICF_{(2)}^\prime,\label{identities}
\end{align}
where $s_{IJ}$ is an arbitrary symmetric tensor and 
\begin{align}
F_{(2)}=\frac{\nabla_\mu p_\nu}{4(h^2-1)}p^\lambda\Gamma_\lambda\Gamma_5\Gamma^{\mu\nu},\,\,\,\,\,\,F_{(2)}^\prime=\frac{\nabla_\mu p_\nu}{4(h^2-1)}\Gamma^{\mu\nu}p^\lambda\Gamma_\lambda\Gamma_5.
\end{align}
The last two identities in (\ref{identities}) follows from Killing spinor equation (\ref{KSeq}) and the supersymmetry surviving condition (\ref{defofcalOtil}).

With these identities, it is shown that the following term
\begin{align}
I_1&\equiv\int\!\mathrm{d}^5x\sqrt{g}\mathrm{tr}\Bigg(i\lambda_Ip^\mu(t^{IJ}-\epsilon^{IJ}F_{(2)}^\prime)\Gamma_\mu\lambda_J\Bigg)
\end{align}
transforms as
\begin{align}
\delta_\xi I_1=\int\!\mathrm{d}^5x\sqrt{g}\mathrm{tr}\Bigg(2ip^\mu t^{IJ}D_{IJ}\xi_K\Gamma_\mu\lambda^K+{\cal O}((D_{IJ})^0)\Bigg)
\end{align}
and thus cancels the terms linear in $D_{IJ}$
in $\delta_{\xi} S_{vec}^\prime$.

With this $I_1$ added, 
the unwanted terms are
\begin{align}
&\delta_\xi(S_{\text{vec}}^\prime+I_1)-\int\!\mathrm{d}^5x\sqrt{g}\mathrm{tr}\Bigg(
  4it^{IJ}(\xi_I\Gamma^m\lambda_J{\cal O}(1,1)A_m-\xi_I\lambda_J{\cal O}(1,1)\sigma)\Bigg)\nonumber\\
=&\int\!\mathrm{d}^5x\sqrt{g}\mathrm{tr}\Bigg(
  \frac{1}{2}\left(2i(\nabla^\mu p^\nu)(D_\mu\sigma)\xi_I\Gamma_\nu\lambda^I-i(\nabla^\mu p^\nu)F_{\mu\nu}\xi_I\lambda^I\right)-\frac{6i}{\ell^2}p^\mu A_{t^\prime}\xi_I\Gamma_\mu\lambda^I\nonumber\\
&\phantom{\int\!\mathrm{d}^5x\sqrt{g}\mathrm{tr}\Bigg(}-i\epsilon^{\mu\nu\lambda\rho}(\nabla_\mu p_\nu)F_{\lambda 5}\xi_I\Gamma_\rho\lambda^I+\frac{i}{2}\epsilon^{\mu\nu\lambda\rho}(\nabla_\mu p_\nu)F_{\lambda\rho}\xi_I\Gamma_t\lambda^I\Bigg),
\end{align}
where we arranged all the terms like $\xi_I(\cdots)\lambda_J$ into the form of $(\cdots)_{IJ}\xi_K(\cdots)\lambda^K$ with the help of (\ref{identities}).
Now it is easy to see that 
this is equal to
\begin{align}
\delta_\xi\int\!\mathrm{d}^5x\sqrt{g}\mathrm{tr}\Bigg(\frac{1}{2}\epsilon^{\mu\nu\lambda\rho}(\nabla_\mu p_\nu)F_{\lambda\rho}\sigma-(\nabla^\mu p^\nu)(F_{\mu\nu}A_5+(\partial_5A_\mu)A_\nu)+\frac{3}{\ell^2}(p^\mu A_\mu)^2\Bigg).
\end{align}

Putting above results together we finally obtain a desired action: 
\begin{align}
&S_{\text{vec}}\nonumber\\
=&\frac{1}{g_{\text{YM}}^2}\int\!\frac{\mathrm{d}^5x}{2\pi\beta}\sqrt{g}\mathrm{tr}\Bigg( \frac{1}{2}F_{mn}F^{mn}-(D_m\sigma)(D^m\sigma) -\frac{1}{2}D_{IJ}D^{IJ}+i\lambda_I\Gamma^mD_m\lambda^I\nonumber\\
&\phantom{\frac{1}{{g_{\text{YM}}^\prime}^2}\int\!\mathrm{d}^5x\sqrt{g}\mathrm{tr}\Bigg(}
-\lambda_I[\sigma,\lambda^I]+2A_{t^\prime}t^{IJ}D_{IJ}+\frac{1}{\ell^2}(3A_{t^\prime}^2-2\sigma^2)\nonumber\\
&\phantom{\frac{1}{{g_{\text{YM}}^\prime}^2}\int\!\mathrm{d}^5x\sqrt{g}\mathrm{tr}\Bigg(}+i\lambda_I\left(t^{IJ}p^\mu\Gamma_\mu+\frac{1}{4}\epsilon^{IJ}(\nabla_\mu p_\nu)\Gamma^{\mu\nu}\Gamma_5\right)\lambda_J\nonumber\\
&\phantom{\frac{1}{{g_{\text{YM}}^\prime}^2}\int\!\mathrm{d}^5x\sqrt{g}\mathrm{tr}\Bigg(}-\frac{1}{2}\epsilon^{\mu\nu\lambda\rho}(\nabla_\mu p_\nu)F_{\lambda\rho}\sigma\nonumber\\
&\phantom{\frac{1}{{g_{\text{YM}}^\prime}^2}\int\!\mathrm{d}^5x\sqrt{g}\mathrm{tr}\Bigg(}+(\nabla^\mu p^\nu)(F_{\mu\nu}A_5+(\partial_5A_\mu)A_\nu)-\frac{3}{\ell^2}(p^\mu A_\mu)^2\Bigg),\label{Sveccomp}
\end{align}
%
%
%
which is invariant under the supersymmetry transformation $\delta_\xi$ up to massive Kaluza-Klein modes.
Here we multiplied $\frac{1}{2\pi\beta g_{\text{YM}}^2}$ to the action to obtain a four dimensional action normalized in a familiar way.

The theta term 
\begin{align}
\frac{i\theta_0}{32\pi^2}\int \mathrm{d}^4x^\prime\sqrt{G}\mathrm{tr}\epsilon^{\mu^\prime\nu^\prime\lambda^\prime\rho^\prime}F_{\mu^\prime\nu^\prime}F_{\lambda^\prime\rho^\prime}\label{theta}
\end{align}
is also supersymmetry invariant due to its topological nature
and can be added to the action.

The action (\ref{Sveccomp}) should be also invariant under the gauge
transformation ${\cal G}(\epsilon_g)$ with the gauge transformation
parameter satisfying 
${\cal O}(1,1)\epsilon_g=\partial_{t^\prime}\epsilon_g=0$
in order to obtain a gauge invariant action on the squashed four-sphere
after the dimensional reduction.
We can see that this is indeed the case.
Actually, with some computations, we can show that 
the terms in the last line in (\ref{Sveccomp}) can be rewritten as  
\begin{align}
\frac{1}{g_{\text{YM}}^2}\int\!\frac{\mathrm{d}^5x}{2\pi\beta}\sqrt{g}\mathrm{tr}\left((\nabla^\mu p^\nu)F_{\mu\nu}A_{t^\prime}-\frac{\epsilon^2\ell^2}{4}\frac{1-\frac{r^2}{4\ell^2}}{1+\frac{r^2}{4\ell^2}}\epsilon^{\mu\nu\lambda\rho}F_{\mu\nu}F_{\lambda\rho}\right).\label{thetalike}
\end{align}
The second term looks like the
theta term, but it has the non-constant pre-factor.
As we will see in the next section, only the values of the action
around the north pole and the south pole are relevant to the partition
function.
There, the second term in (\ref{thetalike}) behaves like the ordinal theta term.
Note, however, that the non-constant pre-factor $\frac{1-\frac{r^2}{4\ell^2}}{1+\frac{r^2}{4\ell^2}}$ flip its sign as one goes from the north pole to the south pole.
Therefore the contribution of this term cannot be absorbed to the addition of the theta term (\ref{theta}).
Rather, it changes the Yang-Mills coupling constants for the instantons
which are point-wisely localized at the north pole and the south
pole.


\section{Localization}
In this section we will calculate the partition function of the four
dimensional theory on the squashed four-sphere defined in the previous
sections by using the localization technique.

The localization technique can be applied if the theory has a continuous symmetry with a fermionic generator $Q$, for which there exist a fermionic potential $V$ such that
the bosonic part of $S_\text{r}\equiv QV$ is positive semi definite and $QS_{\text{r}}=0$.
Below, we will 
call $V$ as regulator potential and $S_\text{r}$ as regulator action.

To explain the localization technique, we consider following quantity
\begin{align}
Z^\prime (\tau)=\int\!{\cal D}\Phi\mathrm{exp}\left[-S-\tau S_\text{r}\right],
\end{align}
where $\tau$ is a real parameter.
We denote all the fields by $\Phi$.
This $Z^\prime(\tau)$ includes the original partition function as $Z=Z^\prime (0)$.
If $\tau$ is large, all configurations but those around the zeroes of $S_\text{r}$ is suppressed, and in the limit of $\tau\rightarrow\infty$ the saddle point approximation becomes exact:
\begin{align}
Z^\prime(\infty)=\sum_{\Phi_0}\mathrm{exp}[-S[\Phi_0]]\times Z_{\text{1-loop}}(\Phi_0).
\end{align}
where $\Phi_0$ is a zero point configuration of $S_\text{r}$ and $Z_{\text{1-loop}}(\Phi_0)$ is the perturbative one loop determinant of the theory described by the action $S_\text{r}$, around $\Phi=\Phi_0$.
On the other hand, by differentiating $Z^\prime$ by $\tau$ one obtains
\begin{align}
\frac{\mathrm{d}Z^\prime(\tau)}{\mathrm{d}\tau}&=\int{\cal D}\Phi(-S_{\text{r}})\exp[-S-\tau S_{\text{r}}]\nonumber\\
&=\int{\cal D}\Phi Q\left(-V\exp[-S-\tau S_{\text{r}}]\right)\nonumber\\
&=0,
\end{align}
which shows that
$Z^\prime(\tau)$ is actually independent of the value of $\tau$.
Therefore the partition function is
\begin{align}
Z=Z^\prime(\infty)=\sum_{\Phi_0}\mathrm{exp}[-S[\Phi_0]]\times Z_{\text{1-loop}}(\Phi_0).
\end{align}

The 1-loop factor $Z_{\text{1-loop}}(\Phi_0)$ is 
a non-trivial, but, an exactly calculable quantity.
We can decompose $\Phi$ into its bosonic part $\Phi_b$ and fermionic part $\Phi_f$, and then can chose $V$ as
\begin{align}
V=((Q\Phi_f)^\dagger,\Phi_f),
\end{align}
where $^\dagger$ is an Hermitian conjugate and $(\cdot,\cdot)$ is a $Q^2$ invariant positive definite inner product of functions.
$S_{\text{r}}=QV$ with this $V$ indeed satisfies the required properties for the localization.
If we decompose $\Phi_b$ and $\Phi_f$ further as $\Phi_b=(X, Q\Xi)$ and $\Phi_f=(QX, \Xi)$, fix Hermicity of fields appropriately and write the regulator potential $V$ as
\begin{align}
V=
\begin{bmatrix}
QX& \Xi
\end{bmatrix}
\begin{bmatrix}
D_{0,0}&D_{0,1}\\
D_{1,0}&D_{1,1}
\end{bmatrix}
\begin{bmatrix}
X\\
Q\Xi
\end{bmatrix},
\end{align}
then the regulator action $S_\text{r}$ is
\begin{align}
S_r=
\begin{bmatrix}
X& Q\Xi
\end{bmatrix}
\begin{bmatrix}
{\cal H}&0\\
0&1
\end{bmatrix}
&
\begin{bmatrix}
D_{0,0}&D_{0,1}\\
D_{1,0}&D_{1,1}
\end{bmatrix}
\begin{bmatrix}
X\\
Q\Xi
\end{bmatrix}\nonumber\\
&+
\begin{bmatrix}
QX& \Xi
\end{bmatrix}
\begin{bmatrix}
D_{0,0}&D_{0,1}\\
D_{1,0}&D_{1,1}
\end{bmatrix}
\begin{bmatrix}
1&0\\
0&{\cal H}
\end{bmatrix}
\begin{bmatrix}
QX\\
\Xi
\end{bmatrix},\label{Sr}
\end{align}
where
\begin{align}
{\cal H}\equiv Q^2.
\end{align}
From (\ref{Sr}) it immediately follows that
\begin{align}
Z_{\text{1-loop}}=\left(\frac{\mathrm{det}_{\mathrm{coKer}(D_{1,0})}{\cal H}}{\mathrm{det}_{\mathrm{Ker}(D_{1,0})}{\cal H}}\right)^{\frac{1}{2}}.
\end{align}
If the eigenvalues of ${\cal H}$ are $h_i$ and their degeneracies in $\mathrm{Ker}(D_{1,0})$ and in $ \mathrm{coKer}(D_{1,0})$ are $n_{\text{b}i}$ and $n_{\text{f}i}$ respectively, the 1-loop determinant can be written as
\begin{align}
Z_{\text{1-loop}}=\prod_ih_i^{\frac{n_{\text{f}i}-n_{\text{b}i}}{2}}.
\end{align}
Here 
$n_{\text{b}i}$ and $n_{\text{f}i}$
can be read off from the index of $D_{1,0}$ defined by
\begin{align}
\mathrm{ind}(D_{1,0},{\cal H}; q)\equiv\mathrm{tr}_{\mathrm{coKer}(D_{1,0})}\mathrm{e}^{q{\cal H}}-\mathrm{tr}_{\mathrm{Ker}(D_{1,0})}\mathrm{e}^{q{\cal H}}.\label{index}
\end{align}
Note that this index is well defined only when both of $n_{\text{b}i}$ and $n_{\text{f}i}$ are finite for every eigenvalue.
This is satisfied when $D_{1,0}$ is transversally elliptic, which have been checked in many cases \cite{Hama:2012bg,Pestun:2007rz}.

In this section we first consider the 5d theory on the twisted $S^4\times S^1$ with the twisted periodicity (\ref{twper}) and obtain the saddle point configurations and 1-loop determinants.
Then 
we consider only 
the Kaluza-Klein zero modes which satisfy (\ref{twistedrestriction}) 
and obtain the
results on the squashed four-sphere.
After this dimensional reduction, 
there are the supersymmetry invariant actions (\ref{Shypcomp}) and
(\ref{Sveccomp}), thus we obtain the partition function by evaluating
them at the saddle point configurations.

Although the saddle point configurations and 1-loop determinants are already obtained in the 5d theory on the untwisted $S^4\times S^1$ in \cite{Kim:2012gu,Terashima:2012ra}, it is not straightforward to obtain our result from them.
First, the saddle point configurations obtained in the untwisted theory may be inconsistent to the twisted periodicity.
Second, since the periodicity in our theory is different from that in the untwisted theory and they give different eigenvalues for the same differential operator, the 1-loop determinant must be changed even if the saddle point configurations are same.\footnote{
One may also worry that the transversal ellipticity of $D_{1,0}$ may be violated because of the twisted periodicity.
However, it is not the case, since the notion of transversal
ellipticity is invariant under any continuous (small) deformation of the
differential operator.
}

As we will see in the following, the saddle point configurations are the same.
On the other hand, the 1-loop determinants and the values of the action
at the saddle point configurations are different from those in the
untwisted theory, which makes the partition function non-trivial.

%
%
%
%
%
%

\subsection*{The vectormultiplet}
In our case we can chose $Q$ as supersymmetry $\delta_\xi$ with some fixed killing spinor $\xi_I$.
We take $\xi_I$ bosonic to have $\delta_\xi$ fermionic.
In this section, we denote this fermionic $\delta_\xi$ as $\delta$.

We chose the $\xi_I$ or, equivalently, $\psi_I$ as
\begin{align}
\Gamma^{\hat{1}\hat{2}}\psi_2=\Gamma^{\hat{3}\hat{4}}\psi_2=i\psi_2,
\end{align}
so that the supersymmetry generated by this $\xi_I$ remains in the four dimensional limit, and normalize them as
\begin{align}
s&=\xi_I\xi^I=-\mathrm{cos}\theta,\nonumber\\
v&=\xi_I\Gamma^m\xi^I\partial_m=\frac{i}{\ell}\left(-x_2\frac{\partial}{\partial x_1}+x_1\frac{\partial}{\partial x_2}-x_4\frac{\partial}{\partial x_3}+x_3\frac{\partial}{\partial x_4}\right)+\frac{\partial}{\partial t}.
\end{align}

As discussed in the beginning of this section, $V_{\text{vec}}$, regulator potential for the vectormultiplet, can be chosen as\footnote{
One should consider the regulator potential for the hypermultiplet
simultaneously because the original action is invariant only when both
the vectormultiplet and the hypermultiplet are simultaneously
transformed by (\ref{deltalambda}) and (\ref{deltahyp}).
As we will see later, 
however, the regulator potential for the hypermultiplet neither affects
the saddle point condition of the vectormultiplet nor includes
fluctuations in the vectormultiplet around the saddle point
configurations.
Therefore one can forget the hypermultiplet completely when one calculate the contribution to the 1-loop determinant from the vectormultiplet.
}
\begin{align}
V_{\text{vec}}=\int\!\mathrm{d}^5x\sqrt{\mathrm{det}g}\delta\left(\mathrm{tr}(\lambda_I)^\dagger\lambda_I\right),
\end{align}
where
\begin{align}
(\delta\lambda_I)^\dagger
  =\frac{1}{2}\xi_I^\dagger\Gamma^{mn}F_{mn}
  -\xi_I^\dagger\Gamma^mD_m\sigma
  +\xi_J^\dagger D_J^{\phantom{J}I}
  -2\xi_K^\dagger\Gamma_5t_K^{\phantom{K}I}\sigma.\label{deltalambdadagger}
\end{align}
(\ref{deltalambdadagger}) follows from (\ref{deltalambda}) and Hermicity condition of fields
\begin{align}
\sigma^\dagger=-\sigma,\,\,\,\,\,\,(D_{IJ})^\dagger=-D^{IJ}
\end{align}
which are required for the convergence of the original path integral $Z$.

The bosonic part of $\delta V_{\text{vec}}$ is obviously positive.
Concretely, it is
\begin{align}
&\delta V_{\text{vec}}|_{\text{bos}}\nonumber\\
=&\int\!\mathrm{d}^5x\sqrt{g}\mathrm{tr}\Bigg(\frac{1}{2}F_{mn}F^{mn}-(D_m\sigma)(D^m\sigma)-\frac{1}{2}D_{IJ}D^{IJ}-\frac{1}{\ell^2}\sigma^2\nonumber\\
&\phantom{\int\!\mathrm{d}^5x\sqrt{g}\mathrm{tr}\Bigg(}-\frac{s}{4}\epsilon^{\mu\nu\lambda\rho}F_{\mu\nu}F_{\lambda\rho}-\frac{s}{4}\epsilon^{\mu\nu\lambda\rho}F_{\mu\nu}F_{\lambda\rho}-\frac{1}{2}\epsilon^{\mu\nu\lambda\rho}F_{\mu\nu}(D_\lambda v_\rho)\sigma\Bigg)\nonumber\\
=&\int\!\mathrm{d}^5x\sqrt{g}\mathrm{tr}\Bigg(
  \frac{1-s}{2}\left((F_+)_{\mu\nu}-\frac{\sigma}{2(1-s)}((dv)_+)_{\mu\nu}\right)^2\nonumber\\
&\phantom{\int\!\mathrm{d}^5x\sqrt{g}\mathrm{tr}\Bigg(}+\frac{1+s}{2}\left((F_-)_{\mu\nu}+\frac{\sigma}{2(1+s)}((dv)_-)_{\mu\nu}\right)^2+F_{\mu t}F^{\mu t}\nonumber\\
&\phantom{\int\!\mathrm{d}^5x\sqrt{g}\mathrm{tr}\Bigg(}-(D_m\sigma)(D^m\sigma)-\frac{1}{2}D_{IJ}D^{IJ}\Bigg),\label{SLvecbos}
\end{align}
where $(dv)_{\mu\nu}=\partial_\mu v_\nu-\partial_\nu v_\mu$ and the subscript $\pm$ denotes the sd/asd decomposition of a two-form defined as
\begin{align}
(F_\pm)_{\mu\nu}\equiv \frac{1}{2}\left(F_{\mu\nu}\pm\frac{1}{2}\epsilon_{\mu\nu}^{\phantom{\mu\nu}\lambda\rho}F_{\lambda\rho}\right).
\end{align}

Now we investigate the zero configurations of (\ref{SLvecbos}).
When $s\neq \pm 1$, all the squares must vanish at a saddle point:
\begin{align}
&F_{\mu t}=0,\,\,\,\,\,\,D_m\sigma=0,\,\,\,\,\,\,D_{IJ}=0\nonumber\\
&F_+-\frac{\sigma}{2(1-s)}(dv)_+=0,\,\,\,\,\,\,F_-+\frac{\sigma}{2(1+s)}(dv)_-=0,\label{vecsaddleeq}
\end{align}
which implies
\begin{align}
A_\mu=0,\,\,\,\,\,\,A_{t^\prime}=a_{t^\prime},\,\,\,\,\,\,\sigma=0,\,\,\,\,\,\,D_{IJ}=0\label{gfsaddle}
\end{align}
up to a gauge transformation.
Here $a_{t^\prime}$ is a constant in the Lie algebra, which is periodic because of the large gauge transformations in $t^\prime$ unfixed by (\ref{gfsaddle}):
\begin{align}
a_{t^\prime}\sim a_{t^\prime}+\frac{n_iH_i}{\beta}\,\,\,\,\,\,(n_i\in \mathbb{Z}),
\end{align}
where $H_i$ are the basis of the Cartan subalgebra of the gauge group.

At the point $s=\pm 1$ (north/south pole), on the other hand, one of the last two conditions in (\ref{vecsaddleeq}) is not imposed since the coefficient of corresponding term in (\ref{SLvecbos}) vanishes.
Thus instanton configurations which are pointwisely localized at the two poles are also allowed as saddle point configurations.

Evaluating the Yang-Mills action (\ref{Sveccomp}) 
with the theta term (\ref{theta}) 
at these saddle points,
we have
\begin{align}
S_{\text{vec}}=\frac{8\pi^2\ell^2}{g_{\text{YM}}^2}\mathrm{tr}a_{t^\prime}^2+\frac{16\pi^2}{{g^\prime_{\text{YM}}}^2}(-\nu_{\text{np}}+\nu_{\text{sp}})+2i\theta_0(\nu_{\text{np}}+\nu_{\text{sp}}),\label{vecsaddlevalue}
\end{align}
where
\begin{align}
\frac{1}{{g^\prime_{\text{YM}}}^2}=\frac{1+\epsilon^2\ell^2}{g_{\text{YM}}^2}.\label{gYMeff}
\end{align}
The first term comes from the mass term of $A_{t^\prime}$ in the second line of (\ref{Sveccomp}).
The second term and the third term are the contributions of the instantons which are pointwisely localized at the north pole or the south pole.
At the north pole $(s=-1)$, $F_-$ are unconstrained and thus there are anti-instantons.
We denoted the instanton number as $\nu_{\text{np}}$, which is non-positive integer.
At the south pole $(s=1)$, on the other hand, the self-dual instanton solutions are allowed, the instanton number of which we denote as $\nu_{\text{sp}}\ge 0$.
Because of the second term in (\ref{thetalike}), the Yang-Mills coupling
constant in the instanton contributions is effectively
changed.\footnote{
Here the Yang-Mills coupling constant itself is not changed
because, for example, the local high energy scattering of the gluons
are described by $g_{\text{YM}}$, not $g'_{\text{YM}}$.
Only the weights for the instanton contributions are changed by the term.
}

To evaluate the saddle point
value of the action,
we restricted the two squashing parameters such that
$\epsilon_1=-\epsilon_2\equiv\epsilon$.
It may be hard task, but interesting to find the Yang-Mills action 
even for $\epsilon_1 \neq -\epsilon_2$
although there would not exist such term.

\subsubsection*{BRST complex}
To calculate the 1-loop determinant of the vectormultiplet, here we introduce ghosts which consist of the fermionic ones $c,\tilde{c}, c_0, \tilde{c}_0$, and the bosonic ones $B, B_0, a_0, \tilde{a}_0$.
We define the BRST symmetry $\delta_B$, under which the ghosts transform as
\begin{align}
\delta_Bc&=icc+a_0,\,\,\,\,\,\, \delta_Ba_0=0,\,\,\,\,\,\, \delta_B\tilde{c}=B,\,\,\,\,\,\, \delta_BB=i[a_0,\tilde{c}],\nonumber\\
\delta_B\tilde{a}_0&=\tilde{c}_0,\,\,\,\,\,\,\delta_B\tilde{c}_0=i[a_0,\tilde{a}_0],\,\,\,\,\,\, \delta_BB_0=c_0,\,\,\,\,\,\, \delta_Bc_0=i[a_0,B_0]
\end{align}
and all the other fields in the vectormultiplet and the hypermultiplet transform as
\begin{align}
\delta_B(\text{boson})={\cal G}(c)(\text{boson}),\,\,\,\,\,\,\delta_B(\text{fermion})=i\{c,(\text{fermion})\}.
\end{align}
We define the supersymmetry transformation law of ghosts as
\begin{align}
\delta c=-i(s\sigma-v^mA_m),\,\,\,\,\,\,\delta B=iv^m\partial_m\tilde{c},\,\,\,\,\,\, \delta (\text{other ghosts})=0
\end{align}
so that the square of the new fermionic charge defined by $Q=\delta+\delta_B$ is the same for all the fields:
\begin{align}
{\cal H}\equiv Q^2=i{\cal L}(v)-2it_I^{\phantom{I}J}+{\cal G}(a_0).\label{calH5d}
\end{align}
%

To fix the gauge, we introduce a $Q$-exact term $S_{\text{GF}}=QV_{\text{GF}}$ with
\begin{align}
V_{\text{GF}}=\int\!\mathrm{d}^5x\sqrt{g}\mathrm{tr}(\tilde{c}(G+B_0)+c\tilde{a}_0),
\end{align}
where
\begin{align}
G=i\nabla^mA_m+i{\cal L}(v)(\Phi-A_5)=i\nabla^\mu A_\mu+iv^\mu\partial_\mu(v^\nu A_\nu-s\sigma)+i\partial_5\Phi
\end{align}
with $\Phi=v^mA_m-s\sigma$.
One can show by explicit calculation that this term play the same role as $\delta_B V_{\text{GF}}$ and fix the gauge properly after the ghosts being integrated out. 

Now $\Phi=\{A_m,\sigma,D_{IJ},\lambda_I\}\oplus \{\text{ghosts}\}$.
Then we replace $V_{\text{vec}}$ with
\begin{align}
V^\prime_{\text{vec}}=V_{\text{vec}}+V_{\text{gho}}+V_{\text{GF}},
\end{align}
where $V_{\text{gho}}=\int\!\mathrm{d}^5x\sqrt{g}\mathrm{tr}((Qc)^\dagger c+(Q\tilde{c})^\dagger \tilde{c})$.
The saddle point equations are the previous ones (\ref{vecsaddleeq}) from $QV_{\text{vec}}|_{\text{bos}}=0$ and
\begin{align}
-i(s\sigma-v^mA_m)+a_0=0,\,\,\,\,\,\,B=0
\end{align}
from $QV_{\text{gho}}|_{\text{bos}}=0$.

For systematic calculation, it is convenient to replace the spinor $\lambda_I$ with the fields with integral spins \cite{Terashima:2012ra}:
\begin{align}
\Psi\equiv\delta\sigma=i\xi_I\lambda^I,\,\,\,\,\,\,\Psi_\mu\equiv\delta A_\mu=i\xi_I\Gamma_\mu\lambda^I,\,\,\,\,\,\,\Xi_{IJ}\equiv\xi_I\Gamma_5\lambda_J+\xi_J\Gamma_5\lambda_I.
\end{align}
These can be solved for $\lambda_I$ as
\begin{align}
\lambda_I=-i\Gamma_5\xi_I\Psi-i\Gamma^{\mu 5}\xi_I\Psi_\mu+\xi^J\Xi_{JI}.
\end{align}
With these, we divide the set of fields $\Phi$ into 
\begin{align}
X&=(\sigma,A_\mu,\bar{a}_0,B_0),\,\,\,\,\,\,\Xi=(\Xi_{IJ},\bar{c},c)\label{XandXi}
\end{align}
and their supersymmetry partners $QX$ and $Q\Xi$.

Below we will calculate the index (\ref{index}).\footnote{
The 1-loop determinant with the same twisted periodicity was already
calculated in \cite{Kim:2012gu} to compute the superconformal index in
five dimension, although our main concern in this paper is the 
theory on the squashed four-sphere.
The computations in \cite{Kim:2012gu}
are essentially same as the ones in this paper,
however, we use the pure spinor like method for the hypermultiplets
as in \cite{HST, Pestun:2007rz} and construct explicitly 
the BRST complex.
}
Since the gauge transformation in ${\cal H}$ acts uniformly to all the fields, its contribution is factored out in the index:
\begin{align}
\mathrm{ind}(D_{1,0},{\cal H};q)=\sum_{\alpha\in\mathfrak{g}}\mathrm{e}^{iqa_0\cdot \alpha}(-2+\mathrm{ind}^\prime(D_{1,0},{\cal H}(a_0=0);q)),
\end{align}
where $\mathfrak{g}$ is the Lie algebra of the gauge group.
$-2$ is the contribution from two bosonic zero modes $\tilde{a}_0$ and $B_0$.
Here the second term in the parentheses is defined for the remaining
fields, $X^\prime=(A_\mu,\sigma)$ and $\Xi$, which can be further
decomposed into the sum of the contributions from the modes who have the
same Kaluza-Klein momentum $\frac{k}{\beta}$, $X^{\prime(k)}$ and
$\Xi^{(k)}$.
With these, we find
\begin{align}
\mathrm{ind}^\prime(D_{1,0},{\cal H}(a_0=0);q)=\sum_{k\in\mathbb{Z}}\mathrm{ind}^{\prime(k)}(D^{(k)}_{1,0},{\cal H}^{(k)}(a_0=0);q),
\end{align}
where $D^{(k)}_{1,0}$ and ${\cal H}^{(k)}$ are defined by replacing $\partial_{t^\prime}$ with $\frac{ik}{\beta}+\tilde{\cal O}(1,1)$ in $D_{1,0}$ and ${\cal H}$, respectively (see (\ref{twistedKKdec})).
More explicitly, ${\cal H}^{(k)}$ is given by
\begin{align}
{\cal H}^{(k)}=-{\cal L}(v_{\cal H})-i(2t_I^{\phantom{I}J}+\tilde{R}(1,1)_I^{\phantom{I}J})-\frac{k}{\beta}+{\cal G}(a_0),
\end{align}
where
\begin{align}
v_{\cal H}=-i(v^\mu\partial_\mu+p)=\frac{\gamma_i}{\ell}\partial_{\alpha_i}
\end{align}
with 
\begin{eqnarray}
\gamma_i=1-i\ell\epsilon_i,
\end{eqnarray}
and $\tilde{R}$ was defined in (\ref{rt}).
Again, since $-\frac{k}{\beta}$ in ${\cal H}^{(k)}$ acts uniformly to all the fields in $X^{\prime(k)}$ and $\Xi^{(k)}$, its contribution is factored out and thus the index is
\begin{align}
\mathrm{ind}(D_{1,0},{\cal H};q)=\sum_{\alpha\in\mathfrak{g}}\mathrm{e}^{iqa_0\cdot\alpha}\left(-2+\sum_{k\in\mathbb{Z}}\mathrm{e}^{-\frac{qk}{\beta}}\mathrm{ind}^{\prime(0)}(D^{(0)}_{1,0},{\cal H}^{(0)}(a_0=0);q))\right).\label{vindfinal}
\end{align}

In order for the index $\mathrm{ind}^{\prime(0)}$ to be well defined, 
the degeneracy of each eigenvalue of ${\cal H}^{(0)}(a_0=0)$ must be finite both in $\mathrm{Ker}(D^{(0)}_{1,0})$ and in $\mathrm{coKer}(D^{(0)}_{1,0})$.
The sufficient condition for this is that $D^{(0)}_{1,0}$ is a transversally elliptic operator with respect to $v_{\cal H}$.
In our case this is satisfied as discussed in the beginning of this section.

Then $\mathrm{ind}^{\prime(0)}$ can be calculated by applying Atiyah-Bott formula \cite{Atiyah:1984px} to the complex $E_{X^{\prime(0)}}\rightarrow E_{\Xi^{(0)}}$ (where $\Gamma(E_{X^{\prime(0)}})=\{X^{\prime(0)}\}$ and $\Gamma(E_{\Xi^{(0)}})=\{\Xi^{(0)}\}$):
\begin{align}
\mathrm{ind}^{\prime(0)}(D^{(0)}_{1,0},{\cal H}^{(0)}(a_0=0);q)=\sum_{x_p\in F}\frac{\mathrm{tr}_{E_{\Xi^{(0)}}}\mathrm{e}^{q\hat{\cal H}^{(0)}(a_0=0)}-\mathrm{tr}_{E_{X^{\prime(0)}}}\mathrm{e}^{q\hat{\cal H}^{(0)}(a_0=0)}}{\mathrm{det}(1-\frac{\partial x^\prime}{\partial x})}\Big|_{x_p},
\end{align}
where $F$ is the set of the fixed points under $x^\mu\rightarrow x^{\prime\mu}=\mathrm{e}^{-q{\cal L}(v_{\cal H})}x^\mu$, i.e. the north pole and the south pole.
The determinant in the denominator is
\begin{align}
\mathrm{det}\left(1-\frac{\partial x^\prime}{\partial x}\right)=
(1-\mathrm{e}^{\frac{iq\gamma_1}{\ell}})(1-\mathrm{e}^{-\frac{iq\gamma_1}{\ell}})(1-\mathrm{e}^{\frac{iq\gamma_2}{\ell}})(1-\mathrm{e}^{-\frac{iq\gamma_2}{\ell}}).
\end{align}
$\hat{\cal H}^{(0)}$ is the vector bundle homomorphism naturally induced from ${\cal H}^{(0)}$.
The eigenvalues of ${\cal H}^{(0)}(a_0=0)$ are
\begin{align}
\mathrm{e}^{\frac{iq\gamma_1}{\ell}},\mathrm{e}^{-\frac{iq\gamma_1}{\ell}},
\mathrm{e}^{-\frac{iq\gamma_2}{\ell}},\mathrm{e}^{-\frac{iq\gamma_2}{\ell}},1
\end{align}
in $E_{X^{\prime(0)}}$ and
\begin{align}
1,\mathrm{e}^{\frac{iq(\gamma_1+\gamma_2)}{\ell}}
,\mathrm{e}^{-\frac{iq(\gamma_1+\gamma_2)}{\ell}}
,1,1
\end{align}
in $E_{\Xi^{(0)}}$.
Thus
\begin{align}
&\mathrm{ind}^{\prime(0)}(D^{(0)}_{1,0},{\cal H}^{(0)}(a_0=0);q)\nonumber\\
=&\frac{1+\mathrm{e}^{-i\frac{q}{\ell}(\gamma_1+\gamma_2)}}{(1-\mathrm{e}^{-i\frac{q}{\ell}\gamma_1})(1-\mathrm{e}^{-i\frac{q}{\ell}\gamma_2})}+\frac{1+\mathrm{e}^{-i\frac{q}{\ell}(\gamma_1+\gamma_2)}}{(1-\mathrm{e}^{-i\frac{q}{\ell}\gamma_1})(1-\mathrm{e}^{-i\frac{q}{\ell}\gamma_2})}\nonumber\\
=&\sum_{n_1,n_2\ge 0}\left( \mathrm{e}^{-i\frac{q}{\ell}n_i\gamma_i} +\mathrm{e}^{-i\frac{q}{\ell}(n_i+1)\gamma_i} +\mathrm{e}^{i\frac{q}{\ell}n_i\gamma_i} +\mathrm{e}^{i\frac{q}{\ell}(n_i+1)\gamma_i} \right),\label{vindprimefinal}
\end{align}
where we expanded the denominator of the terms coming from the north pole in positive power of $\mathrm{e}^{-i\frac{q}{\ell}\gamma_i}$ and those from the south pole in negative power in the second line, to obtain the correct index \cite{Pestun:2007rz}.
%
%
%
%

{}From (\ref{vindfinal}) and (\ref{vindprimefinal}), we find that 
the 1-loop determinant is
\begin{align}
Z^{\text{vec,5d}}_{\text{1-loop}}&=\prod_{\alpha\in\Delta}(ia_0\cdot \alpha)^{-1}\nonumber\\
&\phantom{=}
\times\prod_{k\in\mathbb{Z}}\prod_{n_1,n_2\ge 0}\left(\frac{k}{\beta}-\frac{in_i\gamma_i}{\ell}+ia_0\cdot\alpha\right)^{\frac{1}{2}}
\left(\frac{k}{\beta}-\frac{i(n_i+1)\gamma_i}{\ell}+ia_0\cdot\alpha\right)^{\frac{1}{2}}\nonumber\\
&\phantom{=\prod_{\alpha\in\Delta}}
\left(\frac{k}{\beta}+\frac{in_i\gamma_i}{\ell}+ia_0\cdot\alpha\right)^{\frac{1}{2}}
\left(\frac{k}{\beta}+\frac{i(n_i+1)\gamma_i}{\ell}+ia_0\cdot\alpha\right)^{\frac{1}{2}}\nonumber\\
&=\prod_{\alpha\in\Delta_+}(ia_0\cdot \alpha)^{-2}\nonumber\\
&\phantom{=}
\times\prod_{k\in\mathbb{Z}}\prod_{n_1,n_2\ge 0}
\left(\frac{k}{\beta}-\frac{in_i\gamma_i}{\ell}+ia_0\cdot\alpha\right)
\left(\frac{k}{\beta}-\frac{i(n_i+1)\gamma_i}{\ell}+ia_0\cdot\alpha\right)\nonumber\\
&\phantom{=\prod_{\alpha\in\Delta}}
\left(\frac{k}{\beta}+\frac{in_i\gamma_i}{\ell}+ia_0\cdot\alpha\right)
\left(\frac{k}{\beta}+\frac{i(n_i+1)\gamma_i}{\ell}+ia_0\cdot\alpha\right),
\end{align}
where $\Delta$ is the set of roots and $\Delta_+$ is the set of positive roots in $\mathfrak{g}$.
We neglected the contribution from the modes in Cartan subalgebra, which gives only an $a_0$ independent overall factor.

Omitting the contributions from the massive Kaluza-Klein modes, we obtain the four dimensional result
\begin{align}
Z^{\text{vec}}_{\text{1-loop}}&=\prod_{\alpha\in\Delta_+}(ia_0\cdot \alpha)^{-2}\prod_{n_1,n_2\ge 0}
\left(-\frac{in_i\gamma_i}{\ell}+ia_0\cdot\alpha\right)
\left(-\frac{i(n_i+1)\gamma_i}{\ell}+ia_0\cdot\alpha\right)\nonumber\\
&\phantom{=\prod_{\alpha\in\Delta_+}(ia_0\cdot \alpha)^{-2}\prod_{n_1,n_2\ge 0}}
\times\left(\frac{in_i\gamma_i}{\ell}+ia_0\cdot\alpha\right)
\left(\frac{i(n_i+1)\gamma_i}{\ell}+ia_0\cdot\alpha\right)\nonumber\\
&=\prod_{\alpha\in\Delta_+}\frac{\Upsilon_b(i\hat{a}_0\cdot\alpha)\Upsilon_b(-i\hat{a}_0\cdot\alpha)}{(\hat{a}_0\cdot\alpha)^2},\label{vec1loop}
\end{align}
where
\begin{align}
b=\sqrt{\frac{\gamma_1}{\gamma_2}}=\sqrt{\frac{1-i\ell\epsilon_1}{1-i\ell\epsilon_2}}
 \, ,\label{defofb}
\end{align}
and 
\begin{eqnarray}
\hat{a}_0=\frac{i\ell}{\sqrt{\gamma_1\gamma_2}}a_0.
\end{eqnarray}
In the last step in (\ref{vec1loop}) we again neglected some $a_0$ independent overall factors.
Here we defined the Upsilon function by the following infinite product
\begin{align}
\Upsilon_b(x)=\prod_{n_1,n_2\ge 0}\left(bn_1+\frac{n_2}{b}+x\right)\left(bn_1+\frac{n_2}{b}+b+\frac{1}{b}-x\right).
\end{align}
$\Upsilon_b(x)$ is also characterized by following relations
\begin{align}
\Upsilon_b(x)&=\Upsilon_b\left(b+\frac{1}{b}-x\right),\,\,\,\,\,\,\Upsilon_b\left(\frac{1}{2}\left(b+\frac{1}{b}\right)\right)=1,\nonumber\\
\Upsilon_b(x+b)&=\Upsilon_b(x)\frac{\Gamma(bx)}{\Gamma(1-bx)}b^{1-2bx},\,\,\,\,\,\,\Upsilon_b\left(x+\frac{1}{b}\right)=\Upsilon_b(x)\frac{\Gamma(\frac{x}{b})}{\Gamma(1-\frac{x}{b})}\left(\frac{1}{b}\right)^{1-\frac{2x}{b}}.
\end{align}

The 1-loop determinant (\ref{vec1loop}) is identical to the result obtained for
the ellipsoid, except that the parameter $b$, which is in the case of
the ellipsoid the square root of the ratio between the length of the
major semi-axis and that of the minor semi-axis and thus must be real,
is replaced by (\ref{defofb}).
$b$ can take arbitrary value in $\mathbb{C}\backslash \mathbb{R}_+$ with $\mathrm{Re}b>0$.
Moreover, when both of $\epsilon_i$ are pure imaginary and $|\epsilon_i|<\ell$, the value of $b$ becomes real positive and the parameter region obtained in \cite{Hama:2012bg} is reproduced.

\subsection*{The hypermultiplet}
Now we consider the contribution from the hypermultiplet.
The regulator potential for the hypermultiplet can be taken as
\begin{align}
V_{\text{hyp}}=\int\!\mathrm{d}^5x\sqrt{g}\mathrm{tr}((\delta\psi)^\dagger\psi).
\end{align}

Here we divide $\psi$ into the components proportional to $\xi$, 
that is, $\delta q_I$, and the components proportional to  $\check{\xi}$ as
\begin{align}
\psi=-\frac{i}{s}(\xi_I\psi_+^I+\check{\xi}_{I^\prime}\psi_-^{I^\prime}),\label{hypredefinition}
\end{align}
where
\begin{align}
\psi_{+I}=-2i\xi_I\psi=\delta q_I,\,\,\,\,\,\,\psi_{-I^\prime}=\check{\xi}_{I^\prime}\psi.
\end{align}
This field redefinition eliminates spinor indices and thus simplify the form of $\delta^2$. 
Since (\ref{hypredefinition}) is an orthogonal decomposition, the saddle
point condition of $S_\text{r}|_{\text{bos}}=Q
V_{\text{hyp}}|_{\text{bos}}$, that is, $\delta\psi=0$, is equivalent to
$\delta\psi_{+I}=\delta\psi_{-I^\prime}=0$, which implies
\begin{align}
q_I={\cal F}_{I^\prime}=0,
\end{align}
with an appropriate Hermicity conditions for $q,F$ as shown 
in \cite{Terashima:2012ra}.
Note that we do not obtain any further constraint on the saddle point configurations of the vectormultiplet, as commented in the footnote when we calculated the contribution of the vectormultiplet.

To continue, we take $\check{\xi}_{I^\prime}$ explicitly as
\begin{align}
\check{\xi}_{I^\prime}=\frac{M^{-\frac{1}{2}}}{2}\left(1-\frac{1}{s}v^n\Gamma_n\right)\eta_{I^\prime},
\end{align}
where $\eta_1^\prime$ and $\eta_2^\prime$ are the solutions of (\ref{KSexplicit}) with the constant spinor chosen as $\Gamma^{\hat{1}\hat{2}}\psi^\prime_2=-\Gamma^{\hat{3}\hat{4}}\psi^\prime_2=i\psi^\prime_2$ and normalized as $\eta_{I^\prime}\eta^{I^\prime}=-s$.
$M$ is a scalar function to normalize $\check{\xi}$ as (\ref{xicheckcond}) which we take as independent of $t$.
This $\check{\xi}_{I^\prime}$ satisfy the following differential equation
\begin{align}
{\cal L}(\partial_{t^\prime})\check{\xi}_{I^\prime}=i(\gamma_1-\gamma_2)t_{I^\prime}^{\phantom{I^\prime}J^\prime}\check{\xi}_{J^\prime}.
\end{align}
where $t_{I^\prime}^{\phantom{I^\prime}J^\prime}$ have the same components as $t_I^{\phantom{I}J}$.
This means that one have to twist the periodicity of the fields further with the rotation of $I^\prime=1,2$:
\begin{align}
\Phi(x^\prime,t^\prime+2\pi\beta)=\mathrm{e}^{-2\pi\beta(\tilde{\cal O}(1,1)-i(\gamma_1-\gamma_2)t_{I^\prime}^{\phantom{I^\prime}J^\prime})}\Phi(x^\prime,t^\prime)\label{twperhyp}
\end{align}
to preserve the supersymmetry on the squashed four-sphere.

To calculate the 1-loop determinant, we choose $X$ as $q_I$ and $\Xi$ as $\psi_{I^\prime}$.
Then the index (\ref{index}) can be calculated just in the same way as in the case of the vectormultiplet:
\begin{align}
\mathrm{ind}(D_{1,0},{\cal H};q) & = 
\sum_{\mathrm{r}\in R}\mathrm{e}^{iqa_0\cdot \mathrm{r}}
\mathrm{ind}^{(k)}(D^{(k)}_{1,0},{\cal H}^{(k)}(a_0=0);q) \nonumber \\
&= \sum_{\mathrm{r} \in R} \mathrm{e}^{iqa_0\cdot \mathrm{r}} 
\sum_{k\in\mathbb{Z}}\mathrm{e}^{-\frac{qk}{\beta}}
\mathrm{ind}^{(0)}(D^{(0)}_{1,0},{\cal H}^{(0)}(a_0=0);q) .
\label{hindfinal}
\end{align}
where ${\cal H}^{(k)}$ is defined by the action of ${\cal H}$ on the fields satisfying the twisted periodic boundary condition (\ref{twperhyp}) with Kaluza-Klein momentum $\frac{k}{\beta}$:
\begin{align}
{\cal H}^{(k)}=-{\cal L}(v_{\cal H})-i(\gamma_1+\gamma_2)t_I^{\phantom{I}J}+({R^\prime}_{I^\prime}^{\phantom{I^\prime}J^\prime}-i(\gamma_1-\gamma_2)t_{I^\prime}^{\phantom{I^\prime}J^\prime})-\frac{k}{\beta}+{\cal G}(a_0)+m.
\end{align}
We can show that $R^\prime=0$ at the north and the south poles.
Here we also added a mass term in the manner as explained in Section 2.
$D_{1,0}^{(k)}$ is defined by the similar restriction, and it is guaranteed by the discussion in the beginning of this section that it is transversally elliptic.
We can also compute $\mathrm{ind}^{(0)}$ in the same way as for the
vector multiplets.
The result is
\begin{align}
\mathrm{ind}^{(0)}=\sum_{n_1,n_2\ge 0}\left(\mathrm{e}^{-i\frac{q}{\ell}\left(n_i+\frac{1}{2}\right)\gamma_i}+\mathrm{e}^{i\frac{q}{\ell}\left(n_i+\frac{1}{2}\right)\gamma_i}\right),\label{hindprimefinal}
\end{align}
which gives the 1-loop determinant for the hypermultiplet on the twisted
$S^4 \times S^1$.

The 1-loop determinant for the hypermultiplet on the squashed four-sphere,
up to irrelevant overall factors,
is obtained by dropping the contributions from massive Kaluza-Klein
modes as
\begin{align}
Z^{\text{hyp}}_{\text{1-loop}}&=\prod_{\mathrm{r}\in R}\prod_{n_1,n_2\ge 0}
\left(i\frac{\gamma_i}{\ell}\left(n_i+\frac{1}{2}\right)+ia_0\cdot\mathrm{r}+m\right)^{-\frac{1}{2}}\nonumber\\
&\phantom{\prod_{\mathrm{r}\in R}\prod_{n_1,n_2\ge 0}}
\left(-i\frac{\gamma_i}{\ell}\left(n_i+\frac{1}{2}\right)+ia_0\cdot\mathrm{r}+m\right)^{-\frac{1}{2}}\nonumber\\
&=\prod_{\mathrm{r}\in R}\Upsilon_b\left(\frac{1}{2}\left(b+\frac{1}{b}\right)+i\hat{a}_0\cdot\mathrm{r}+\frac{i\ell m}{\sqrt{\gamma_1\gamma_2}}\right)^{-\frac{1}{2}}.
\end{align}
where $R$ is the weights of the representation of the hypermultiplet in the gauge group.

\subsection*{Partition function and Wilson loop}

So far we have considered vector saddle point configurations of $F_{\mu\nu}=0$ only.
Relaxing this restriction, however, as we saw in the subsection of the vectormultiplet, we have additional saddle point configurations, in which $F_-\neq 0$ on the north pole and $F_+\neq 0$ on the south pole.
They are weighted by (\ref{vecsaddlevalue}).

The supersymmetry algebras (\ref{vecalg}) at the poles are identical to those of the 4d ${\cal N}=2$ multiplets in the $\Omega$-background, with the parameters of the each theory appropriately identified.
Therefore we have only to quote the result $Z_{\text{inst}}(ia,\varepsilon_1,\varepsilon_2,q)$ in that case, where $a$ is the vev of the Higgs scalar, $(\varepsilon_1,\varepsilon_2)$ are the $\Omega$-deformation parameter and $q=\mathrm{e}^{2\pi i\tau}$ with
\begin{align}
\tau=\tau(g,\theta)=\frac{\theta}{2\pi}+\frac{4\pi i}{g^2}
\end{align}
the complexified coupling constant \cite{Nekrasov:2002qd}.
Here we used the notation adopted in \cite{Pestun:2007rz} for the parameters in $Z_{\text{inst}}$.
With this, the contributions in our case can be written as $|Z_{\text{inst}}\left(-a_0,\frac{\gamma_1}{\ell},\frac{\gamma_2}{\ell},\mathrm{e}^{2\pi i\tau(g_{\text{YM}}^\prime,\theta_0)}\right)|^2$.

Putting all together, the partition function on the squashed four-sphere is
\begin{align}
Z=&\int\mathrm{d}\hat{a}_0\exp\left[-\frac{8\pi^2\gamma_1\gamma_2}{g_{\text{YM}}^2}\mathrm{tr}\hat{a}_0^2\right]
\left|Z_{\text{inst}}\left(-a_0,\frac{\gamma_1}{\ell},\frac{\gamma_2}{\ell},\mathrm{e}^{2\pi i\tau(g^\prime_{\text{YM}},\theta_0)}\right)\right|^2\nonumber\\
&\prod_{\alpha\in\Delta_+}\Upsilon_b(i\hat{a}_0\cdot\alpha)\Upsilon_b(-i\hat{a}_0\cdot\alpha)
\times\prod_{\mathrm{r}\in R}\Upsilon_b\left(\frac{1}{2}\left(b+\frac{1}{b}\right)+i\hat{a}_0\cdot\mathrm{r}+\frac{i\ell m}{\sqrt{\gamma_1\gamma_2}}\right)^{-\frac{1}{2}}.\label{Zfinal}
\end{align}
where $g^\prime_{\text{YM}}$ is defined by (\ref{gYMeff}).

In the similar way, one can also calculate the expectation value of any gauge invariant operator which is invariant under $\delta_\xi$.
One important example is the Wilson loop.
Since the combination
\begin{align}
v^mA_m-s\sigma=i\gamma_1A_{\alpha_1^\prime}+i\gamma_2A_{\alpha_2^\prime}+A_{t^\prime}-s\sigma
\end{align}
is supersymmetry invariant, one can construct Wilson loops which are
both gauge invariant and supersymmetry invariant.
For the generic values of the squashing parameters $\epsilon_i$,
we have two kinds of the supersymmetric Wilson loops which 
are defined by
\begin{align}
W_i(\theta^\prime)&=\mathrm{tr}\mathrm{P}
\exp\left[\ell\int_{L_i}\mathrm{d}  \lambda
\left(iA_{\alpha_i^\prime}+\frac{1}{\gamma_i}(A_{t^\prime}-s\sigma)\right)\right],
\end{align}
where $i=1,2$ and the two loops $L_1$ and $L_2$ are
\begin{align}
L_1=\{(\theta^\prime,\phi^\prime, \alpha_1^\prime,\alpha_2^\prime)
 =(c_1,0,\lambda,c_2) 
|0\le \lambda \le
 2\pi\}, \\
 L_2=\left\{(\theta^\prime,\phi^\prime,\alpha_1^\prime,\alpha_2^\prime)=
\left(c_1,\frac{\pi}{2},c_2,\lambda\right)
|0\le \lambda \le  2\pi\right\},
\end{align}
where $c_1, c_2$ are constants.
The expectation values of these Wilson loops are immediately obtained by inserting the following saddle point values
\begin{align}
W_1(\theta^\prime)=\mathrm{tr}\exp\left[\frac{2\pi i\hat{a}_0}{b}\right],\,\,\,\,\,\, W_2(\theta^\prime)=\mathrm{tr}\exp[2\pi ib\hat{a}_0]
\end{align}
into the integrand in (\ref{Zfinal}).

Though the 1-loop determinant is calculated for general $(\epsilon_1,\epsilon_2)$, the full result (\ref{Zfinal}) is valid only for the special case of $\epsilon_1=-\epsilon_2=\epsilon$ where we constructed the action of the vectormultiplet.
In this case, the partition function (\ref{Zfinal}) takes the following form
\begin{align}
Z(\tau,b,\mu)=&\int\mathrm{d}a\exp\left[-\frac{8\pi^2}{g^2}\mathrm{tr}a^2\right]
\left|Z_{\text{inst}}\left(ia,b,\frac{1}{b},\mathrm{e}^{2\pi i\tau(g,\theta)}\right)\right|^2\nonumber\\
&\prod_{\alpha\in\Delta_+}\Upsilon_b(ia\cdot\alpha)\Upsilon_b(-ia\cdot\alpha)
\times\prod_{\mathrm{r}\in R}\Upsilon_b\left(\frac{1}{2}\left(b+\frac{1}{b}\right)+ia\cdot\mathrm{r}+i\mu\right)^{-\frac{1}{2}},\label{forAGT}
\end{align}
with the substitutions of $g=\frac{g_{\text{YM}}}{\sqrt{\gamma_1\gamma_2}}$, $\theta=\theta_0$, $b=\sqrt{\frac{\gamma_1}{\gamma_2}}$ and $\mu=\frac{\ell m}{\sqrt{\gamma_1\gamma_2}}$. 
On the other hand, the partition function in the case of the ellipsoid is also in this form, with $g=g_{\text{YM}}$, $b=\sqrt{\frac{\ell}{\tilde{\ell}}}$ \cite{Hama:2012bg}.

In this sense the partition function on the squashed four sphere coincide with that on the ellipsoid.
Moreover, $b=\sqrt{\frac{\gamma_1}{\gamma_2}}$ also can be complex with $|b|=1$ as well as real, unlike in the case of the ellipsoid where $b$ is given as the square root of the ratio between the length of the major semi-axis and that of the minor semi-axis and thus always real.  
That is, by considering the squashed four-spheres we can realize more general parameter region than in the case of the ellipsoids.


For $\epsilon_1 \neq -\epsilon_2$,
if there is a corresponding CFT,
the central charge would be complex,
thus, there could not be an AGT-like relation 
and the partition function could be different from 
the function (\ref{forAGT}).

\section*{Acknowledgments}
We thanks to Kazuo Hosomichi and Naofumi Hama for useful discussion.
The works of T.N. is partly supported by the JSPS Research Fellowships for Young Scientists.
S.T. was supported in part by JSPS KAKENHI Grant Number 23740189.


\vspace{1cm}


\begin{thebibliography}{999}
\parskip=-2pt

\bibitem{Pestun:2007rz}
  V.~Pestun,
  ``Localization of gauge theory on a four-sphere and supersymmetric Wilson loops,''
  Commun.\ Math.\ Phys.\  {\bf 313} (2012) 71
  [arXiv:0712.2824 [hep-th]].

\bibitem{Alday:2009aq}
  L.~F.~Alday, D.~Gaiotto and Y.~Tachikawa,
  ``Liouville Correlation Functions from Four-dimensional Gauge Theories,''
  Lett.\ Math.\ Phys.\  {\bf 91} (2010) 167
  [arXiv:0906.3219 [hep-th]].

\bibitem{Hama:2012bg}
  N.~Hama and K.~Hosomichi,
  ``Seiberg-Witten Theories on Ellipsoids,''
  JHEP {\bf 1209} (2012) 033
   [Addendum-ibid.\  {\bf 1210} (2012) 051]
  [arXiv:1206.6359 [hep-th]].

\bibitem{Gadde:2011ia}
  A.~Gadde and W.~Yan,
  ``Reducing the 4d Index to the $S^3$ Partition Function,''
  JHEP {\bf 1212} (2012) 003
  [arXiv:1104.2592 [hep-th]].

\bibitem{Imamura:2011uw}
  Y.~Imamura,
  ``Relation between the 4d superconformal index and the $S^3$ partition function,''
  JHEP {\bf 1109} (2011) 133
  [arXiv:1104.4482 [hep-th]].

\bibitem{Liu:2012bi}
  J.~T.~Liu, L.~A.~Pando Zayas and D.~Reichmann,
  ``Rigid Supersymmetric Backgrounds of Minimal Off-Shell Supergravity,''
  JHEP {\bf 1210} (2012) 034
  [arXiv:1207.2785 [hep-th]].

\bibitem{Imamura:2012bm}
  Y.~Imamura,
  ``Perturbative partition function for squashed $S^5$,''
  arXiv:1210.6308 [hep-th].

\bibitem{Kim:2012gu}
  H.~-C.~Kim, S.~-S.~Kim and K.~Lee,
  ``5-dim Superconformal Index with Enhanced En Global Symmetry,''
  JHEP {\bf 1210} (2012) 142
  [arXiv:1206.6781 [hep-th]].

\bibitem{Terashima:2012ra}
  S.~Terashima,
  ``On Supersymmetric Gauge Theories on $S^4$ x $S^1$,''
  arXiv:1207.2163 [hep-th].

\bibitem{Harlow:2011ny}
  D.~Harlow, J.~Maltz and E.~Witten,
  ``Analytic Continuation of Liouville Theory,''
  JHEP {\bf 1112} (2011) 071
  [arXiv:1108.4417 [hep-th]].

\bibitem{Hama:2011ea}
  N.~Hama, K.~Hosomichi and S.~Lee,
  ``SUSY Gauge Theories on Squashed Three-Spheres,''
  JHEP {\bf 1105} (2011) 014
  [arXiv:1102.4716 [hep-th]].

\bibitem{Imamura:2011wg}
  Y.~Imamura and D.~Yokoyama,
  ``N=2 supersymmetric theories on squashed three-sphere,''
  Phys.\ Rev.\ D {\bf 85} (2012) 025015
  [arXiv:1109.4734 [hep-th]].

\bibitem{Closset:2012ru}
  C.~Closset, T.~T.~Dumitrescu, G.~Festuccia and Z.~Komargodski,
  ``Supersymmetric Field Theories on Three-Manifolds,''
  JHEP {\bf 1305} (2013) 017
  [arXiv:1212.3388].

\bibitem{Alday:2013lba}
  L.~F.~Alday, D.~Martelli, P.~Richmond and J.~Sparks,
  ``Localization on Three-Manifolds,''
  arXiv:1307.6848 [hep-th].

\bibitem{Hosomichisprivatecommunication}
K.~Hosomichi,
private communication.

\bibitem{Closset:2013vra}
  C.~Closset, T.~T.~Dumitrescu, G.~Festuccia and Z.~Komargodski,
  ``The Geometry of Supersymmetric Partition Functions,''
  arXiv:1309.5876 [hep-th].

\bibitem{Pan}
  Y.~Pan,
  ``Rigid Supersymmetry on 5-dimensional Riemannian Manifolds and Contact Geometry,''
  arXiv:1308.1567 [hep-th].


\bibitem{SuTe}
  S.~Sugishita and S.~Terashima,
  ``Exact Results in Supersymmetric Field Theories on Manifolds with Boundaries,''
  arXiv:1308.1973 [hep-th].

\bibitem{HoOk}
  D.~Honda and T.~Okuda,
  ``Exact results for boundaries and domain walls in 2d supersymmetric theories,''
  arXiv:1308.2217 [hep-th].

\bibitem{HoRo}
K.~Hori and M.~Romo,
  ``Exact Results In Two-Dimensional (2,2) Supersymmetric Gauge Theories With Boundary,''
  arXiv:1308.2438 [hep-th].

\bibitem{HST}
  K.~Hosomichi, R.~-K.~Seong and S.~Terashima,
  ``Supersymmetric Gauge Theories on the Five-Sphere,''
  Nucl.\ Phys.\ B {\bf 865} (2012) 376
  [arXiv:1203.0371 [hep-th]].

\bibitem{Atiyah:1984px}
  M.~F.~Atiyah and R.~Bott,
  ``The Moment map and equivariant cohomology,''
  Topology {\bf 23} (1984) 1.

\bibitem{Nekrasov:2002qd}
  N.~A.~Nekrasov,
  ``Seiberg-Witten prepotential from instanton counting,''
  Adv.\ Theor.\ Math.\ Phys.\  {\bf 7} (2004) 831
  [hep-th/0206161].

\end{thebibliography}
\end{document}